\documentclass[a4paper,11pt]{article}
\pdfoutput=1 % if your are submitting a pdflatex (i.e. if you have
\usepackage{jheppub} % for details on the use of the package, please
\usepackage[T1]{fontenc} % if needed

\usepackage{mathtools}
\usepackage{amssymb}
\usepackage{tensor}
\usepackage{braket}
\usepackage{cleveref}
\usepackage{booktabs}
\usepackage{diagbox}
\usepackage{transparent}
\usepackage{xcolor}

\allowdisplaybreaks
\usepackage[normalem]{ulem}

\crefname{table}{table}{tables}
\Crefname{table}{Table}{Tables}
\crefname{section}{section}{sections}
\Crefname{section}{Section}{Sections}
\crefname{figure}{figure}{figures}
\Crefname{figure}{Figure}{Figures}
\crefname{equation}{eq.}{eqs.}
\Crefname{equation}{Eq.}{Eqs.}
\crefname{appendix}{appendix}{appendices}
\Crefname{appendix}{Appendix}{Appendices}

\newcommand{\R}{\mathbb{R}}
\newcommand{\Z}{\mathbb{Z}}
\newcommand{\cA}{\mathcal{A}}
\newcommand{\cF}{\mathcal{F}}
\newcommand{\cL}{\mathcal{L}}
\newcommand{\tr}{\operatorname{tr}}
\newcommand{\sgn}{\operatorname{sgn}}
\newcommand{\dd}{\mathrm{d}}
\newcommand{\vev}[1]{\left\langle #1\right\rangle}
\newcommand{\order}{\mathcal{O}}

\title{
High-quality Axion Strings from Monopole Strings}

\author[a,b]{Nathaniel Craig}
\author[a]{and Amalia Madden}

\affiliation[a]{Kavli Institute for Theoretical Physics, Santa Barbara, CA 93106, USA}
\affiliation[b]{Department of Physics, University of California, Santa Barbara, CA 93106, USA}

\emailAdd{ncraig@ucsb.edu}
\emailAdd{amadden@ucsb.edu}

\abstract{Five-dimensional~'t Hooft--Polyakov monopole strings whose cores restore a non-abelian gauge symmetry become axion strings for extra-dimensional axions. The parametric separation between the string tension and the scale of quantum gravity admits a viable post-inflationary cosmology for high-quality axion dark matter.}

\begin{document}
\maketitle
\flushbottom

\section{Introduction}
The QCD axion is one of the best-motivated extensions of the Standard Model \cite{PhysRevLett.38.1440, PhysRevLett.40.223, PhysRevLett.40.279}. Originally introduced to solve the strong CP problem, it dynamically relaxes the observed QCD theta angle, $\bar{\theta}$, to zero, thereby explaining the apparent smallness of the neutron electric dipole moment.
The axion is also a compelling dark matter candidate: for suitable cosmological histories, its relic abundance can account for all of the observed dark matter \cite{Preskill:1982cy, Abbott:1982af, Dine:1982ah, Davis:1986xc}.
Moreover, axion-like fields are a generic feature of string compactifications, where they descend from the nontrivial topology of the extra dimensions \cite{Svrcek:2006yi, Arvanitaki_2010}.

In conventional four-dimensional field theory, the QCD axion is a pseudo-Nambu--Goldstone boson arising from the spontaneous breaking of an anomalous global Peccei--Quinn symmetry, $U(1)_{\rm PQ}$. Below the PQ-breaking scale, this symmetry is realized nonlinearly as an axion shift symmetry. Non-perturbative QCD effects associated with the anomaly break the continuous shift symmetry and generate the axion potential required to solve the strong CP problem. 

The cosmological history of the axion depends crucially on when PQ symmetry is broken relative to inflation. 
If PQ symmetry is broken before or during inflation, the axion field is correlated across the observable universe, and cosmological production proceeds through the misalignment mechanism: coherent oscillations of the axion field about the minimum of its potential give rise to a cold dark matter density. 
If PQ symmetry is instead broken after inflation, the initial axion angle takes different values in different patches of the universe. In this case, axion strings are produced at the PQ phase transition, while domain walls form later when QCD effects generate the axion potential. The dark matter abundance then receives contributions not only from misalignment, but also from the evolution and decay of this string-wall network.

However, an ordinary four-dimensional global PQ symmetry does not by itself
fully solve the strong CP problem. Since PQ symmetry is anomalous, a
high-quality solution must also explain why there are no additional
shift-symmetry-violating operators in the effective theory that could displace the axion from its
CP-conserving minimum \cite{Kamionkowski:1992mf, PhysRevD.46.539, Holman_1992, Ghigna:1992iv}. This axion quality problem is exacerbated by the expectation that quantum gravity violates global symmetries: even if gravity is the only source of PQ violation, generic Planck-suppressed operators in the low-energy EFT can still be large enough to spoil the axion solution unless they are forbidden to extremely high dimension.

Reference \cite{Lu_2024} emphasized that attempts to improve the quality of an ordinary four-dimensional $U(1)_{\rm PQ}$ axion can come into severe tension with post-inflationary axion cosmology. For example, in models where dangerous PQ-violating operators are forbidden by a discrete $\Z_p$ symmetry, the defects produced at the PQ phase transition need not be the elementary axion strings required for the usual $(N_{\rm DW}=1)$ string-wall network to collapse before it dominates the energy density. More broadly, quality-protected models can introduce additional cosmological obstacles, including stable or long-lived relics, fractionally charged states, and low-energy Landau poles.
Two model-building solutions to the quality problem that preserve standard post-inflationary cosmology were recently proposed in \cite{Petrossian-Byrne:2025mto,Loladze:2025uvf}. In both cases, the axion still arises from a four-dimensional PQ symmetry whose spontaneous breaking produces ordinary axion strings, while extra-dimensional gauge redundancy or higher-form symmetry protects the axion from generic local PQ-breaking operators.

Extra-dimensional axion models \cite{Witten:1984dg,Choi:1985je,Barr:1985hk,Choi:1985bz,
Dine:1986bg,Arkani-Hamed:2003xts,Choi:2003wr,Svrcek:2006yi} provide an alternative and particularly elegant solution to the axion quality problem. In this case, the axion does not
arise as the Goldstone boson of a four-dimensional global $U(1)_{\rm PQ}$
symmetry, but instead as the Wilson line of a higher-dimensional gauge field.
Its shift symmetry is then protected by the one-form global symmetry associated with the higher-dimensional gauge theory \cite{Reece:2023czb, Craig:2024dnl}. Violations of this symmetry typically arise from nonlocal effects around the
compact dimension, making them exponentially suppressed.

At first sight, however, this protection seems to remove the most essential ingredient of post-inflationary axion phenomenology. Without a PQ symmetry, there are no solitonic axion strings whose cores restore the symmetry, much less an obvious production mechanism for
axion strings or domain walls. This is not to say that there are no axion strings whatsoever, but rather that they are typically fundamental objects whose tension lies at or above the scale at which local quantum field theory breaks down \cite{Dolan:2017vmn, Reece:2018zvv, Lanza:2020qmt, Lanza:2021udy, March-Russell:2021zfq, Reece:2024wrn}. While it has been observed \cite{Benabou:2023npn} that solitonic strings for extra-dimensional axions can arise from axion-radion mixing, they still possess Planck-scale tensions and corresponding cosmological challenges (modulo potential production mechanisms in warped compactifications \cite{Cline:2024vbd}).

But there are abundant indications that other solitonic strings may be relevant to extra-dimensional axions. As a periodic scalar, the four-dimensional axion possesses a two-form winding symmetry under which axion string worldsheets are charged. When the axion emerges from an abelian gauge theory in an extra
dimension, this winding symmetry descends from the two-form magnetic symmetry of the
five-dimensional abelian gauge theory. The objects charged under this two-form symmetry are monopole string worldsheets.  In five dimensions, the spontaneous breaking of non-abelian gauge theories can give rise to solitonic 't Hooft--Polyakov monopole strings~\cite{Boyarsky:2002ck,Dermisek:2002ri}, in direct analogy with 't Hooft--Polyakov monopoles~\cite{tHooft:1974kcl,Polyakov:1974ek} in four dimensions. This naturally suggests \cite{Sehayek:2026pvu} that extra-dimensional axion strings are solitonic monopole strings that
survive compactification.

In this paper, we pursue this simple field-theoretic interpretation of extra-dimensional axion strings: they are five-dimensional monopole strings whose cores restore a non-abelian gauge symmetry, rather than a PQ global symmetry. As the relevant scales are all associated with the five-dimensional gauge theory, rather than gravity, this allows for greater parametric scale separation and potentially viable post-inflationary axion cosmology.

This paper is organized as follows: In Sec.~\ref{sec:5d}, we introduce the five-dimensional gauge theory, describe its monopole string solutions, and show how the axion--Yang Mills coupling is generated by an induced Chern--Simons term. In Sec.~\ref{sec:s1}, we study compactification on $S^1$, emphasizing the relation between monopole strings and four-dimensional axion strings. In Sec.~\ref{sec:orbifold-global}, we turn to the phenomenologically relevant $S^1/\Z_2$ orbifold and discuss the corresponding spectrum, axion coupling, quality, and string solutions. In Sec.~\ref{sec:string-cosmology}, we analyze the implications for post-inflationary axion cosmology, including the creation of the string network, radion stabilization, scaling, and dark matter production. We conclude in Sec.~\ref{sec:conclusion}.

\section{Five-dimensional theory}
\label{sec:5d}

We consider UV completions of the axion--Yang Mills EFT in a flat five-dimensional spacetime with coordinates
\begin{equation}
  x^M=(x^\mu,y),\qquad \mu=0,1,2,3
\end{equation}
and two possible compactifications of the fifth dimension: either the circle compactification
\begin{equation}
  y\sim y+L,\qquad L=2\pi R,
\end{equation}
or an orbifold interval,
\begin{equation}
  0\leq y\leq \ell,\qquad \ell=\pi R,
\end{equation}
corresponding to $S^1/\Z_2$ from a covering circle of length $2\ell=2\pi R$. The former compactification is illustrative, while the latter is more phenomenologically relevant. The bulk 5d physics is largely similar, modulo a few subtleties involving the fixed planes of the orbifold. Further conventions are specified in Appendix~\ref{app:setup}. 

In both cases, we take the bulk gauge group to be
\begin{equation}
  G_5=SU(N)\times SU(2) \, ,
\end{equation}
where the $SU(N)$ factor is a proxy for QCD and the $SU(2)$ factor will be spontaneously broken to $U(1)$, which will ultimately give rise to both the four-dimensional axion and corresponding axion strings \cite{Sehayek:2026pvu}. In principle, the spontaneous breaking $SU(2) \rightarrow U(1)$ can be accomplished by the vacuum expectation value (vev) of either a bulk scalar field $\Phi^a$ in the adjoint representation or a Wilson line wrapping the compact dimension via the Hosotani mechanism. In practice, however, the Hosotani mechanism does not allow for sufficient scale separation between the scales of the monopole and the compactification itself, whereas the vev of the bulk scalar provides the necessary parametric freedom; in what follows we will focus on the bulk scalar.  The Wilson line of the surviving $U(1)$ becomes a four-dimensional axion, while the breaking $SU(2) \rightarrow U(1)$ gives rise to 't Hooft--Polyakov monopole strings. Monopole strings whose extended direction is perpendicular to the compact dimension become four-dimensional axion strings whose cores restore the 5d $SU(2)$ gauge symmetry.

The alert reader has no doubt noticed that our bulk gauge theory does not, on its own, admit the sort of mixed Chern--Simons term that would give rise to an axion--Yang Mills coupling in the 4d effective theory. However, the necessary Chern--Simons term may be generated by bulk Dirac fermions $\Psi$ transforming under both $SU(N)$ and $SU(2)$ that acquire the majority of their mass from the vev of $\Phi$. Below the mass of the fermions, the bulk theory exhibits a $U(1) \times SU(N)^2$ Chern--Simons term in order to cancel the anomalies of fermion zero modes living on the monopole string. This delivers the desired axion coupling to $SU(N)$ instanton density in the four-dimensional effective theory. 

The bulk theory must satisfy a few parametric hierarchies of scale in order for this picture to remain consistent. The five-dimensional scale of quantum gravity is 
\begin{equation}
M_{5} = \left( \frac{M_{\rm Pl}^2}{\cL} \right)^{1/3}    
\end{equation}
where $M_{\rm Pl}$ is the 4d Planck scale and $\cL=L,\ell$ for the circle and interval compactifications, respectively. In what follows, the 5d Planck scale lies well above all other scales of interest; field theory prevails.
Gravity aside, a weakly coupled five-dimensional $SU(N)$ gauge theory is necessarily an effective field theory with a cutoff of order
\begin{equation}
  \Lambda_5\sim \frac{24\pi^3}{N g_{5,*}^2},
\label{eq:cutoff}
\end{equation}
where $g_{5,*}$ denotes the largest five-dimensional gauge coupling, coming from the fact that the loop expansion at energy $E$ is controlled by $N g_{5,*}^2 E/(24\pi^3)\ll1$. While the gauge coupling $g_{5,2}$ of the bulk $SU(2)$ is a free parameter set by the desired relationship between the compactification scale and the axion decay constant, the gauge coupling $g_{5,N}$ of the bulk $SU(N)$ would be set by reproducing the QCD gauge coupling in a fully realistic axion model. In terms of the 4d gauge coupling $g_4^2 = g_5^2 / 2 \pi R$, the hierarchy between the cutoff $\Lambda_5$ and $1/R$ is $\Lambda_5 R \sim 12 \pi^2 / N g_4^2$; running the QCD coupling up to typical axion decay constants, this leaves a window between the cutoff and compactification scale of roughly two orders of magnitude. This compressed hierarchy is common to most flat extra-dimensional theories with the Standard Model gauge fields in the bulk, and suffices for our purposes. 

We additionally require the breaking $SU(2) \rightarrow U(1)$ and generation of the CS term to be effectively five-dimensional, which requires the masses of the fermions $\Psi$, the $SU(2)/U(1)$ gauge bosons $W^\pm$, and the radial mode $\rho$ to satisfy $m_\psi R, \, m_W R, \, m_\rho R \gg 1$. The same parametric hierarchies also protect the quality of the resulting axion, since misaligned contributions to the axion potential are exponentially small in these ratios. Finally, the monopole string core should fit well inside the compact direction in order for our description of the 5d topological defects to remain under control. Its size is $\delta_{\rm core}\sim \max(m_W^{-1},m_\rho^{-1}),$
which is already satisfied by the above constraints on the mass spectrum.
     
While the story of extra-dimensional axions from 5d abelian gauge fields is by now familiar, the consequences of a bulk $SU(2)$ embedding are somewhat less so. In what follows, we explore the bulk 5d physics of 't Hooft--Polyakov monopole strings and the induced Chern--Simons term before turning to the consequences of compactification.

\subsection{The 't Hooft--Polyakov monopole string}
\label{sec:noncompact-monostring}

The Higgsed $SU(2)$ theory contains smooth
monopole string solutions, which will ultimately give rise to axion strings in four dimensions. Although the asymptotics and global topology of these string solutions depend on the compactification, the properties of the string cores are insensitive to the compactification provided $m_W R, \, m_\rho R \gg 1$. 

These monopole strings are obtained by lifting the ordinary four-dimensional
't Hooft--Polyakov monopole~\cite{tHooft:1974kcl,Polyakov:1974ek} to five dimensions~\cite{Boyarsky:2002ck, Dermisek:2002ri}. Let the string extend along a spatial direction $s$. The fields depend on the three
transverse coordinates $x^i$, $i=1,2,3$. In the BPS limit, the static equations reduce to
\begin{equation}
  B_i^a=D_i\Phi^a,
  \qquad
  B_i^a=\frac12\epsilon_{ijk}F^a_{jk}.
\end{equation}
With $r=\sqrt{x^ix^i}$, the charge-one solution is
\begin{equation}
  \Phi^a=v_5 H(\xi)\hat x^a,
  \qquad
  A_i^a=\frac{1}{g_{5,2}}\epsilon_{aij}\hat x^j\frac{1-K(\xi)}{r},
  \qquad
  \xi=m_W r,
\end{equation}
where the Prasad-Sommerfield profiles are~\cite{Prasad:1975kr,Bogomolny:1975de}
\begin{equation}
  K(\xi)=\frac{\xi}{\sinh\xi},
  \qquad
  H(\xi)=\coth\xi-\frac1\xi.
\end{equation}
The BPS tension is
\begin{equation}
  T_M^{\rm BPS}=\frac{4\pi v_5}{g_{5,2}}
  =\frac{4\pi m_W}{g_{5,2}^2}.
\end{equation}
Away from the BPS limit the radial profiles obey second-order ODEs and must be found
numerically, but the parametric result is
\begin{equation}
  T_M=\frac{4\pi m_W}{g_{5,2}^2}B(m_\rho/m_W),
  \qquad
  B(0)=1,
  \label{eq:monostring-tension}
\end{equation}
where $B$ is an order-one function. The fate of these monopole string solutions in four dimensions depends on the details of the compactification and the orientation of the strings relative to the compact dimension.

\subsection{The induced Chern--Simons term}
\label{sec:fermion-cs}

The second key ingredient is the $U(1) \times SU(N)^2$ Chern--Simons term. The simplest possibility is to consider a single five-dimensional Dirac fermion transforming as
\begin{equation}
  \Psi\sim ({\bf N},{\bf 2})
\end{equation}
under the bulk $SU(N)\times SU(2)$ gauge symmetry, with Lagrangian density
\begin{equation}
  \cL_\Psi=\bar\Psi i\Gamma^M D_M\Psi-y_5\bar\Psi\Phi^a\sigma^a\Psi.
  \label{eq:yukawa}
\end{equation}
Here the $SU(2)$ gauge generators are $T^a=\sigma^a/2$. Working in a basis where $\vev{\Phi^a}=v_5\delta^{a3}$, the two components of the $SU(2)$ doublet have opposite Yukawa masses $m_\pm=\pm y_5v_5$. We use the unit-doublet-charge normalization of the unbroken abelian connection, defined in Appendix \ref{app:setup}, so that these same components have abelian charges $q_\pm=\pm1$.

The charges of the Dirac fermion are such that the microscopic bulk theory need not have any local Chern--Simons terms. Neither $SU(2)^3$ nor $SU(2) \times SU(N)^2$ Chern--Simons terms are allowed, while we have the freedom of choosing local counterterms to set the coefficient of an allowed $SU(N)^3$ Chern--Simons term to zero. However, following \cite{Boyarsky:2002ck}, once $\langle \Phi \rangle$ breaks $SU(2) \rightarrow U(1)$ and gives mass to $\Psi$, bulk $U(1)^3$, $U(1)$-gravity, and $U(1) \times SU(N)^2$ Chern--Simons terms will be generated in order to cancel the anomalies of fermion zero modes living on the monopole strings via anomaly inflow.\footnote{As noted in \cite{Boyarsky:2002ck}, which studied the case of a bulk $SU(2)$ gauge symmetry, the $U(1)^3$ and $U(1)$-gravity CS terms are needed to cancel the tangent- and normal-bundle anomalies of the worldsheet zero modes. In our case of a bulk $SU(2) \times SU(N)$ gauge group, the zero modes organize into a single complex left-mover in the $\mathbf{N}$ of $SU(N)$, whose anomaly must be canceled by the bulk $U(1) \times SU(N)^2$ CS term.} In what follows, we will be primarily interested in the $U(1) \times SU(N)^2$ CS term, though the $U(1)^3$ CS term may be phenomenologically interesting for the circle compactification.  

A massive five-dimensional Dirac fermion of abelian charge $q$, mass $m$, and $SU(N)$
representation $R$ induces a parity-odd mixed Chern--Simons term,
\begin{equation}
  \Delta S_{\rm CS}
  =\frac{\Delta k}{8\pi^2}\int \cA\wedge \tr_{\bf N}G\wedge G,
\end{equation}
where $\mathcal{A}$ is the abelian connection, and our conventions are specified in Appendix~\ref{app:setup}. The coefficient $\Delta k$ is given by \cite{Bonetti:2013ela}
\begin{equation}
  \Delta k=-\frac12 q C_R\,\sgn(m),
  \qquad
  \tr_R G^2=C_R\tr_{\bf N}G^2.
  \label{eq:fermion-shift}
\end{equation}

For the single $({\bf N},{\bf 2})$ Dirac fermion with Dynkin index $C_{\bf N}=1$, $\Delta k=-\sgn(y_5v_5)$. Crucially, the two contributions from the $SU(2)$ components of the Dirac fermion do not cancel, as both their charges and signed masses are opposite. This is precisely the bulk CS term required to cancel the $SU(N)$ anomaly from colored zero modes on the monopole string.

Five-dimensional parity in the bulk theory can forbid a bare Dirac mass $M$ for the fermion, but it is briefly worth asking what happens if a bulk mass is included. In this case, the masses are split,
\begin{equation}
  m_\pm=M\pm y_5v_5.
\end{equation}
A nonzero Chern--Simons coefficient requires the two components of the doublet to have opposite-sign masses. If all massive components have the
same sign of mass, the opposite charges cancel in the anomaly. Otherwise, they contribute as before:
\begin{equation}
  \Delta k=\begin{cases}
    -\sgn(y_5v_5), & |M|<|y_5v_5|,\\[4pt]
    0, & |M|>|y_5v_5|.
  \end{cases}
\end{equation}
This is consistent with the physical reason for the appearance of the bulk CS term: when $|M|<|y_5v_5|$ a CS term is still required to cancel anomalies from weakly delocalized fermion zero modes on monopole strings, while for $|M|>|y_5v_5|$ the strings no longer support anomalous zero modes \cite{Bagherian:2023jxy}. Thus as long as the modest condition $|M| < |y_5 v_5|$ is satisfied, the bulk Dirac fermion generates the desired mixed Chern--Simons term between the massless $U(1)$ gauge field and the $SU(N)$ gauge fields.

As is often the case, attention must be paid to the cosmological consequences of these fermions in a fully realistic theory. In the simple model presented here, the fermions would be stable colored relics subject to both cosmological and experimental bounds if abundant in the early universe. Reheating above the scale of $SU(2)$ breaking (as explored in Sec.~\ref{sec:string-cosmology}) would necessarily lead to an appreciable cosmological abundance unless the fermions possess a large bare mass parametrically above the reheating temperature and the phase transition is first-order. Decaying the fermions into Standard Model fields through irrelevant operators requires additional $SU(2)$ multiplets beyond those presented here.

\section{Compactification on $S^1$}
\label{sec:s1}

We now compactify the fifth dimension, starting with the simpler circle compactification. While less realistic than the orbifold due to the proliferation of light states, it provides a clean setting for understanding the essential consequences of compactification.

\subsection{Local spectrum}
Under compactification the low-energy abelian connection (below the scale of $SU(2) \to U(1)$) decomposes as
\begin{equation}
  \cA_M\longrightarrow (\cA_\mu,\cA_5).
\end{equation}
The circle compactification preserves a massless 4d vector zero mode $\cA_\mu^{(0)}$ and a Wilson-line axion from the fifth component of $\cA$. On the circle, this takes the form
\begin{equation}
  \theta(x)=\int_0^L\dd y\,\cA_5(x,y),
  \qquad
  \theta\sim\theta+2\pi.
  \label{eq:theta-def}
\end{equation}
The periodicity is generated by large gauge transformations acting on fields of unit abelian charge. 

The 5d $SU(N)$ gauge field gives rise to both a 4d $SU(N)$ zero mode $G_\mu^{(0)}$ and a four-dimensional adjoint scalar
$G_5^{(0)}$. The would-be zero modes of the massive $SU(2)/U(1)$ gauge bosons $W^\pm$, the radial mode $\rho$, and the heavy fermions $\psi_\pm$ are lifted by their 5d mass terms and accompanied by KK towers. 

As for the Kaluza-Klein (KK) excitations, on $S^1$ a bosonic bulk field admits the decomposition
\begin{equation}
  X(x,y)=\frac{1}{\sqrt L}\sum_{n\in\Z}X_n(x)e^{iny/R},
  \qquad
  m_n^2=m_0^2+\frac{n^2}{R^2}.
\end{equation}
The decomposition for bulk fermions is qualitatively similar but somewhat more involved, and is reserved for Appendix \ref{app:fermion-bcs}. As discussed below in Sec.~\ref{sec:circle-quality}, the KK spectrum of fields charged under the unbroken $U(1)$ accumulate additional $\theta$-dependence.  The schematic
zero-mode and KK spectrum is given in Table~\ref{tab:spectrum}.

\begin{table}
    \centering
\begin{tabular}{c|c}
\toprule
Five-dimensional field & Four-dimensional spectrum on $S^1$\\
\midrule
$SU(N)$ gauge field $G_M$ & $G_\mu^{(0)}$, adjoint scalar $G_5^{(0)}$, KK tower\\
$U(1)$ gauge field $\cA_M$ & $\cA_\mu^{(0)}$, axion $\theta$, KK tower\\
$SU(2)/U(1)$ gauge field $W_M^\pm$ & $m_{W,n}^2\simeq m_W^2+n^2/R^2$\\
Scalar radial mode & $m_{\rho,n}^2\simeq m_\rho^2+n^2/R^2$\\
Heavy fermions $\psi_\pm$ & $m_{\psi,n}^2\simeq m_\psi^2+n^2/R^2$\\
\bottomrule
\end{tabular}
 \caption{Schematic zero-mode and KK spectrum on the circle after $SU(2)\to U(1)$.}
\label{tab:spectrum}
\end{table}

\subsection{Axion--Yang Mills coupling}

The desired axion coupling to the $SU(N)$ zero modes arises automatically from the bulk Chern--Simons term discussed in Sec.~\ref{sec:fermion-cs}. Dimensional reduction of the 5d $U(1)$ kinetic term gives the 4d axion kinetic term
\begin{align}
  S_{\rm kin}
  &=\frac12\int \dd^4x\,f_{S^1}^2(\partial_\mu\theta)^2,
\end{align}
with
\begin{equation}
  f_{S^1}^2=\frac{1}{g_{5,A}^2L}
  =\frac{1}{g_{4,A}^2L^2},
  \qquad
  g_{4,A}^2=\frac{g_{5,A}^2}{L}.
  \label{eq:fa}
\end{equation}
The Chern--Simons term reduces to
\begin{equation}
  \frac{k}{8\pi^2}\int_{M_4\times S^1}\cA\wedge\tr G\wedge G
  \longrightarrow
  \frac{k}{8\pi^2}\int_{M_4}\theta\,\tr G\wedge G.
  \label{eq:axion-coupling-form}
\end{equation}
In terms of the canonically normalized 4d axion field $a\equiv f_{S^1}\theta$, this gives the standard axion-QCD coupling
\begin{equation}
  \cL_{4d}\supset
  \frac{k a}{32\pi^2 f_{S^1}}G^A_{\mu\nu}\widetilde G^{A\mu\nu}
\end{equation}
as desired. In the UV, it was required to cancel the $SU(N)$ gauge anomaly induced by colored zero modes living on monopole strings. In the IR, it cancels the $SU(N)$ gauge anomaly of colored zero modes living on monopole strings perpendicular to the compact dimension.

With $\theta\sim\theta+2\pi$, the corresponding circle domain-wall number is
\begin{equation}
  N_{\rm DW}^{S^1}=|k|.
\end{equation}
For the minimal bulk fermion content above, $|k|=1$ on the circle, so the induced
nonabelian instanton potential has a unique vacuum after accounting for the axion
periodicity.

Thus we see that at energies well below $R^{-1}$, $m_W$, $m_\rho$, and $m_\psi$, the theory of the local fields from circle compactification can be mapped to a four-dimensional axion-Yang Mills EFT plus extra light fields:
the $U(1)$ photon zero mode and the $SU(N)$ adjoint scalar $G_5^{(0)}$. 

\subsection{Axion quality} \label{sec:circle-quality}

Axion quality requires non-QCD contributions to the Wilson-line potential to be small compared to the QCD contribution,
\begin{equation}
  m_a^2f_a^2\simeq \chi_{\rm QCD}\simeq (75\,{\rm MeV})^4,
\end{equation}
more precisely
\begin{equation}
  \Delta V_{\rm UV}\lesssim 10^{-10}\chi_{\rm QCD}.
\end{equation}

The high quality of extra-dimensional axions can be clearly understood from the electric one-form global symmetry of the 5d theory \cite{Craig:2024dnl}. In our model, the emergence of the bulk 5d $U(1)$ gauge symmetry from spontaneous breaking of $SU(2)$ implies that the corresponding continuous electric one-form symmetry protecting the axion is only emergent below the scale $m_W$. Indeed, both the massive $W^\pm$ bosons and fermions are charged under the bulk $U(1)$, explicitly breaking the electric one-form symmetry above the scale of their masses. On the circle compactification, charged particle worldlines wrapping the $S^1$ give rise to a potential for the axion that is exponentially small in $m_W R, m_\psi R$. In this respect, our field-theoretic model for extra-dimensional axion strings possesses mandatory (albeit exponentially small) violations of axion quality.

This can be seen at the level of the compactified effective theory by noting that the KK masses of fields charged under the bulk $U(1)$ are Wilson-line dependent; for a field of abelian charge $q$, the mass is shifted by
\begin{equation}
  \frac{n}{R}\longrightarrow \frac{n}{R}+\frac{q\theta}{L}.
\end{equation}
This applies both to the heavy fermions with $q=\pm1$ and to the massive $W^\pm$ bosons with $q=\pm2$. Integrating out these massive KK modes generates contributions of the form
\begin{equation}
  \Delta V(\theta)\sim -\frac{(m R)^2}{L^4} e^{-mL}\cos(q\theta) \, .
\end{equation}

Clearly axion quality is preserved for $m L \gg 1$.  The upper end is limited by the five-dimensional cutoff. With the QCD coupling setting the cutoff, high-scale one-loop QCD running and NDA give the window
\begin{equation}
  1\ll m L \lesssim \frac{2 \pi^2}{\alpha_s}.
\end{equation}
The irreducible quality violations from bulk fermions and $W^\pm$ bosons are nontrivial; adequate quality requires $mL$ at the upper end of the allowed range.

\subsection{Monopole string parallel to the circle}
\label{sec:S1-parallel}

We next consider the fate of the 5d monopole strings, which depends on their orientation relative to the compact dimension. A monopole string parallel to $S^1$ has worldsheet $(t,y)$, while the transverse space is the noncompact $\R^3$ of the four-dimensional observer. This
case has an exact static solution obtained by taking the ordinary monopole solution of
Sec.~\ref{sec:noncompact-monostring} and making it independent of $y$; periodicity on $S^1$ is automatic.

The resulting four-dimensional object is a magnetic monopole for the four-dimensional $U(1)$.
Its mass is
\begin{equation}
  M_{\rm mon}=T_M L
  =\frac{4\pi m_W L}{g_{5,2}^2}B(m_\rho/m_W)
  =\frac{4\pi m_W}{g_{4,2}^2}B(m_\rho/m_W).
  \label{eq:wrapped-monopole-mass}
\end{equation}
At distances $r\gg R,\delta_{\rm core}$, the magnetic field is the ordinary
four-dimensional zero-mode field of the unbroken $U(1)$,
\begin{equation}
  B_r\simeq \frac{g_m}{4\pi r^2}.
\end{equation}
In the normalization where the minimal electric charge is one,
\begin{equation}
  \int_{S^2_\infty}\cF=2\pi n.
\end{equation}
The colored zero modes living on monopole strings in the noncompact 5d theory become a tower of fermionic modes bound to the 4d monopole.

The 4d monopole is, in some sense, the zero mode of the wrapped monopole string. But a wrapped monopole string can also carry momentum along $S^1$, leading to effective Kaluza-Klein excitations of the monopole. The field-theory construction of such BPS traveling waves and their compactified rings
was studied by Ref.~\cite{BlancoPillado:2006sr}. 

In the thin-string approximation
we can choose worldsheet coordinates
\begin{equation}
  X^0=\tau,
  \qquad
  y=\sigma,
  \qquad
  0\leq \sigma<L.
\end{equation}
A chiral transverse traveling wave has
\begin{equation}
  X^1+iX^2=\rho\,e^{in(\sigma-\tau)/R},
  \qquad
  X^3=0.
  \label{eq:ring-wave}
\end{equation}
At fixed four-dimensional time, the core traces a ring of radius $\rho$. In a Nambu-Goto
approximation,
\begin{equation}
  E=T_ML\left(1+\frac{n^2\rho^2}{R^2}\right),
  \qquad
  P_y=T_ML\frac{n^2\rho^2}{R^2}.
\end{equation}
Quantization of compact momentum $P_y=\frac{N}{R}$ gives a radius
\begin{equation}
  \rho^2=\frac{N}{2\pi T_M n^2}.
  \label{eq:ring-radius}
\end{equation}
Thus excited strings wrapping the $S^1$ give rise to four-dimensional macroscopic rings with $\rho\gg R$ provided
\begin{equation}
  N\gg 2\pi T_Mn^2R^2
  =\frac{4\pi B(m_\rho/m_W)}{g_{4,2}^2}n^2m_WR.
\end{equation}
While such rings are not solutions to the four-dimensional equations of motion, they are supported by their momentum in the fifth dimension, somewhat analogous to superconducting rings supported by their currents. They source both an axion field and a magnetic field of the unbroken $U(1)$. From the axion perspective, they are closed axion string loops supporting a dipolar axion field at long distances, while from the $U(1)$ perspective they support a magnetic monopole field. The phenomenology of these rings is novel and largely unexplored. Although it is tempting to identify these rings with conventional four-dimensional axion strings, they are not the strings we're looking for.

\subsection{Monopole string perpendicular to the circle}
\label{sec:S1-perpendicular}

Now take the monopole string to extend along a noncompact spatial direction, say $z$. This is the case which gives a four-dimensional axion string. The transverse space is
\begin{equation}
  M_\perp=\R^2_{\rho,\varphi}\times S^1_y.
\end{equation}
At large $\rho$, the nonabelian fields have relaxed to the abelian Higgs vacuum, and the
only massless field in the transverse problem is the abelian connection. The Wilson-line
axion is
\begin{equation}
  \theta(\varphi)=\int_0^L\dd y\,\cA_y(\rho,\varphi,y).
\end{equation}
A winding-$n$ axion string obeys
\begin{equation}
  \oint_{S^1_\varphi}\dd\theta=2\pi n.
  \label{eq:axion-winding}
\end{equation}
Equivalently,
\begin{equation}
  \int_{S^1_\varphi\times S^1_y}\cF=2\pi n.
  \label{eq:flux-torus}
\end{equation}
A convenient large-distance gauge representative is
\begin{equation}
  \cA_y\simeq \frac{n}{L}\varphi,
  \qquad
  \cA_\varphi\simeq0,
\end{equation}
understood patchwise, with a large gauge transformation across the branch cut in $\varphi$. Then
\begin{equation}
  \cF_{\varphi y}\simeq \frac{n}{L}.
\end{equation}
The radial magnetic field in the transverse
$\R^2\times S^1$ is
\begin{equation}
  B_\rho\simeq \frac{\cF_{\varphi y}}{\rho}=\frac{n}{L\rho}.
\end{equation}
The tail energy per unit four-dimensional length is
\begin{align}
  \mu_{\rm tail}
  &=\frac{1}{2g_{5,A}^2}\int \rho\dd\rho\dd\varphi\dd y\,
  \frac{\cF_{\varphi y}^2}{\rho^2}
  \notag\\
  &=\pi n^2 f_{S^1}^2\log\frac{\rho_{\rm IR}}{\rho_{\rm UV}}.
  \label{eq:tail-tension}
\end{align}
Thus at distances $\rho\gg R$ the object is precisely a four-dimensional global axion
string. The core, however, is not a fundamental PQ scalar core; it is the smooth nonabelian
monopole-string core in which the five-dimensional $SU(2)$ is restored.

As for the compact direction, the perpendicular circle problem is not solved by simply imposing periodicity on the
spherically symmetric monopole. The transverse space is $\R^2\times S^1$, not $\R^3$.
The ordinary $O(3)$ symmetry is lost, and the full Yang-Mills-Higgs equations become
nonlinear elliptic PDEs in $(\rho,y)$, with nontrivial angular dependence carried by the
axion winding. The situation is reminiscent of a related problem, namely periodic monopoles on $\R^2\times S^1$.
The Bogomolny equations on $\R^2\times S^1$ have no
nontrivial finite-energy solutions with ordinary finite-energy boundary conditions; the
interesting periodic monopoles instead have Higgs fields growing logarithmically at
infinity and are related by a Nahm transform to Hitchin equations on a cylinder
\cite{Cherkis:2000cj}. Explicit
closed-form solutions are generally unavailable; the Nahm construction is implemented
numerically even for the simplest chains~\cite{Ward:2005pm}.

Our physical problem is not exactly the BPS periodic-monopole problem, since we want the
Higgs magnitude to approach its fixed vacuum value,  $|\Phi|\to v_5,$
whereas the BPS equation would require a logarithmically varying Higgs field if the
magnetic field has a $1/\rho$ tail. Therefore the compactified axion-string solution with
fixed $|\Phi|\to v_5$ is intrinsically non-BPS in its far field. Although we have been unable to find an elementary analytic solution, the existence and properties of the solution are strongly controlled in our regime of validity $ m_W R, \, m_\rho R \gg 1$. It is useful to study the string in three regions:

\paragraph{Region I: nonabelian core.}
Near the core, at distances
\begin{equation}
  r_{3d}=\sqrt{\rho^2+(y-y_0)^2}\ll R,
\end{equation}
the compact dimension is locally invisible. The massive core profiles are exponentially
close to those of the ordinary noncompact monopole string. The abelian magnetic tail,
which is massless in the Higgs phase, receives compact-image corrections that are
power-suppressed in the local expansion near the core when $\delta_{\rm core}\ll R$.

\paragraph{Region II: local abelian Coulomb regime.}
For
\begin{equation}
  \delta_{\rm core}\ll r_{3d}\ll R,
\end{equation}
the massive $W$ bosons and radial scalar have decayed. The field is abelian and locally
looks like a monopole in $\R^3$:
\begin{equation}
  B\sim \frac{g_m}{4\pi r_{3d}^2},
\end{equation}
where $g_m$ is fixed by the local flux normalization $\int_{S^2} \cF=2\pi n$.

\paragraph{Region III: compactified far field.}
For
\begin{equation}
  \rho\gg R,
\end{equation}
only the $y$-zero mode survives. The field crosses over to
\begin{equation}
  B_\rho\sim \frac{n}{L\rho}.
\end{equation}
This is the global axion-string tail.

Putting it all together, the total tension through an IR scale $\rho_{\rm IR}$ is therefore
\begin{equation}
  \mu_n(\rho_{\rm IR})
  =T_{M,n}+\Delta\mu_{\rm match}
  +\pi n^2f_{S^1}^2\log\frac{\rho_{\rm IR}}{R}+\cdots,
  \label{eq:circle-perp-tension}
\end{equation}
where $T_{M,1}=T_M$. In the BPS multi-monopole limit $T_{M,n}=|n|T_M$, while away from
that limit it is an order-$|n|T_M$ core contribution whose detailed value depends on
short-distance interactions. The logarithmic term is fixed by the axion winding and scales
as $n^2$, so higher-winding global strings generically prefer to split into unit strings when
kinematically allowed.  The contribution from intermediate scales is 
\begin{equation}
  \Delta\mu_{\rm match}\sim \order\left(\frac{1}{g_{5,2}^2R}\right).
\end{equation}
For $m_WR\gg1$, this is parametrically smaller than the core tension,
\begin{equation}
  T_M\sim \frac{m_W}{g_{5,2}^2}\gg \frac{1}{g_{5,2}^2R}.
\end{equation}

These are the axion strings we're looking for, albeit in an axion--Yang Mills EFT with additional vectors, scalars, monopoles, and monopole rings.

\section{Compactification on $S^1/\Z_2$}
\label{sec:orbifold-global}

We next compactify on the orbifold interval $S^1/\Z_2$. Some care must be taken with the $SU(2)$ orbifold global structure, which is invisible in the local KK spectrum but relevant to the properties of the extra-dimensional axion and its strings. In what follows, for concreteness we restrict our attention to conventional orbifolds in which the fields are defined on the covering circle with suitable identifications under reflection. More exotic theories can be defined by exercising the freedom to modify the bulk theory at the orbifold fixed planes.

\subsection{Local spectrum}
The orbifold should be imposed in such a way as to remain compatible with the bulk $SU(2)$ gauge symmetry while dispensing with the $U(1)$ vector zero mode. For the adjoint action we may take $P=\sigma^1$ (corresponding to the $SU(2)$ element $i\sigma^1$, as the overall phase drops out of conjugation). The gauge field is taken to obey
\begin{equation}
  A_\mu(x,-y)=P A_\mu(x,y)P^{-1},
  \qquad
  A_5(x,-y)=-P A_5(x,y)P^{-1}.
  \label{eq:gauge-parity}
\end{equation}
Since
\begin{equation}
  P T^1P^{-1}=+T^1,
  \qquad
  P T^{2,3}P^{-1}=-T^{2,3},
\end{equation}
the vector parities are
\begin{equation}
  A_\mu^1: \text{ even},
  \qquad
  A_\mu^{2,3}: \text{ odd},
\end{equation}
and the scalar fifth components have the opposite parities:
\begin{equation}
  A_5^1: \text{ odd},
  \qquad
  A_5^{2,3}: \text{ even}.
\end{equation}
For the adjoint scalar we choose
\begin{equation}
  \Phi(x,-y)=-P\Phi(x,y)P^{-1}.
  \label{eq:scalar-parity}
\end{equation}
so that $\Phi^1$ is odd and $\Phi^{2,3}$ are even. The vev $\vev{\Phi^a}=v_5\delta^{a3}$
is allowed, which Higgses $SU(2) \rightarrow U(1)$. The corresponding vector
$A_\mu^3$ is odd and has no zero mode, while $A_5^3$ is even and has a zero mode, which will ultimately give rise to our axion. In our normalization conventions, $\cA_5=(g_{5,2}/2)A_5^3$ is the unit-charge Cartan connection.

In contrast, the
component $A_\mu^1$ is even at the orbifold level, but it does not commute with the
$T^3$ vev and obtains a mass of order $m_W$. Similarly, the scalar $A_5^2$ obtains a mass of order $m_W$. Thus no massless abelian vector zero modes or additional scalar zero modes beyond $\cA_5$
survive. 

At this point, it is worth pausing to contemplate the surviving gauge symmetry on the fixed planes. On the orbifold, an infinitesimal gauge transformation parameter $\epsilon$ obeys
\begin{equation}
  \epsilon(x,-y)=P\epsilon(x,y)P^{-1}.
\end{equation}
Thus the infinitesimal gauge transformation in the $T^3$ direction is odd and vanishes at the fixed planes. Globally, however, finite Cartan transformations are more subtle. Writing a
Cartan transformation in unit-doublet-charge normalization as
\begin{equation}
  U(y)=e^{i\lambda(y)\sigma^3},
  \qquad
  \cA\longrightarrow \cA - \dd\lambda,
\end{equation}
the orbifold condition requires $\lambda$ to satisfy
\begin{equation}
  \lambda(-y)=-\lambda(y)\quad \text{mod }2\pi.
  \label{eq:lambda-orbifold-condition}
\end{equation}
At a fixed plane this allows not only $\lambda=0$ but also $\lambda=\pi$ modulo $2\pi$,
corresponding to the $\Z_2$ center element $-{\bf 1}\in SU(2)$. Consequently, a continuous transformation by
\begin{equation}
  \lambda(y)=\frac{\pi y}{\ell},
  \qquad 0\leq y\leq \ell,
  \label{eq:center-large-gauge}
\end{equation}
is therefore an allowed large gauge transformation on the $SU(2)$ orbifold.\footnote{This follows from our strict interpretation of the covering-circle orbifold. One can certainly define interval theories with different boundary global structures, e.g.~treating a center transformation on the boundary as a global symmetry rather than a gauge redundancy. This correspondingly modifies the discussion of axion periodicity and the spectrum of monopole strings.} 

This has significant implications for the axion. On the interval, we might have expected to identify the axion with an $\cA_5$ Wilson line stretching between the two fixed
planes, but under the gauge transformation in Eq.~\ref{eq:center-large-gauge}, 
the half-interval holonomy
\begin{equation}
  \theta_I(x)\equiv\int_0^\ell\dd y\,\cA_5(x,y)
  \label{eq:theta-half-interval}
\end{equation}
 shifts by $\pi$. The $2\pi$-periodic axion is therefore not $\theta_I$, but rather the integral over the covering circle \cite{Hall:2001tn, Haba:2002py},
\begin{equation}
\vartheta(x)=\int_{0}^{2\ell} \dd y \, \cA_5(x,y) =2\int_0^\ell\dd y\,\cA_5(x,y) = 2\theta_I(x),
  \qquad
  \vartheta\sim\vartheta+2\pi.
  \label{eq:vartheta-def}
\end{equation}

As we will see, this has implications for both the induced CS coupling and the monopole string solutions. Although we have arrived at this conclusion for a bulk $SU(2)$ gauge symmetry, similar considerations arise for extra-dimensional axions arising directly from a $U(1)$ gauge theory on $S^1/\Z_2$.

The bulk $SU(N)$ factor is much more straightforward. We choose $SU(N)$ parity assignments so that $G_\mu$ is even and $G_5$ is odd. Thus the low-energy zero-mode spectrum contains only
\begin{equation}
  G_\mu^{(0)},\qquad \vartheta(x)=2\int_0^\ell \dd y\,\cA_5(x,y) \, .
\end{equation}
 There is no $U(1)$ vector zero mode and no $SU(N)$ adjoint scalar zero mode.

As for the KK spectrum, for a bosonic field even under the orbifold,
\begin{equation}
  X_+(x,y)=\frac{1}{\sqrt\ell}X_0(x)+\sqrt{\frac{2}{\ell}}\sum_{n\geq1}X_n(x)\cos\frac{ny}{R},
\end{equation}
whereas for an odd bosonic field,
\begin{equation}
  X_-(x,y)=\sqrt{\frac{2}{\ell}}\sum_{n\geq1}X_n(x)\sin\frac{ny}{R}.
\end{equation}
The KK masses are $n/R=n\pi/\ell$ modulo axion dependence. As with the circle compactification, the decomposition for bulk fermions is somewhat more involved, and is reserved for Appendix \ref{app:fermion-bcs}.

\subsection{Axion--Yang Mills coupling}

The axion coupling to $SU(N)$ zero modes is sensitive to the orbifold and requires some care. Dimensional reduction of the $\cA_5$ kinetic term gives
\begin{equation}
  S_{\rm kin}^{\rm orb}
  =\frac12\int\dd^4x\,f_{\rm orb}^2(\partial_\mu\vartheta)^2,
  \qquad
  f_{\rm orb}^2=\frac{1}{4g_{5,A}^2\ell}.
  \label{eq:forb}
\end{equation}
while local bulk Chern--Simons density on the interval remains
\begin{equation}
  S_{\rm CS}^{I}
  =
  \frac{k}{8\pi^2}\int_{M_4\times I}\cA\wedge\tr_{\bf N}G\wedge G.
  \label{eq:interval-cs}
\end{equation}
Its zero-mode reduction then gives
\begin{equation}
  S_{\rm CS}^{I}\longrightarrow
  \frac{k}{2}\frac{1}{8\pi^2}\int_{M_4}\vartheta\,\tr_{\bf N}G\wedge G.
  \label{eq:interval-cs-reduction}
\end{equation}
At first glance, this poses an apparent puzzle: as we have seen, integrating out a single Dirac fermion generates $|k| = 1$, so that the resulting axion--Yang Mills coupling appears to no longer be properly quantized. Indeed, under $\vartheta\to\vartheta+2\pi$, this local term shifts by $k \pi Q_N$, where
\begin{equation}
  Q_N=\frac{1}{8\pi^2}\int_{M_4}\tr_{\bf N}G\wedge G\in\Z.
\end{equation}

The resolution is familiar \cite{ArkaniHamed:2001is, Scrucca:2001eb, Pilo:2002hu, Barbieri:2002ic}: the local Chern--Simons term on the interval is not the full orbifold fermion determinant. On the circle, the two Cartan components $\psi_\pm$ are independent Kaluza-Klein towers.  On the orbifold, the reflection exchanges them, so they form a single projected tower whose determinant depends both on the bulk fields and on the boundary conditions at the two fixed planes. The parity-odd phase accumulated in the bulk gives the term in Eq.~\ref{eq:interval-cs}, while
the phase associated with the orbifold boundary conditions gives an equal
contribution; together they reproduce the properly-quantized result. This can be seen explicitly by studying the regularized fermion determinant, see Appendix \ref{app:trueCS}. The result is that a single bulk $({\bf N},{\bf 2})$ Dirac fermion in fact delivers the properly quantized 4d coupling
\begin{equation}
  S_{\vartheta GG}
  =
  -\sgn(y_5v_5)
  {1\over8\pi^2}
  \int_{M_4}\vartheta\,\tr_{\bf N}G\wedge G.
\end{equation}

Thus the orbifold compactification is much closer than the circle compactification to a minimal 4d axion--Yang Mills EFT in its local spectrum. As we will see, this also carries over into the spectrum of monopole strings.

\subsection{Axion quality}

Axion quality on the interval is much like axion quality on the circle, up to minor complications associated with the fixed planes. Corrections to the axion potential remain exponentially small; potentially problematic fixed plane-localized terms that could give a polynomially (rather than exponentially) volume-suppressed mass to the axion are forbidden by the bulk gauge symmetry (equivalently, the electric one-form global symmetry) \cite{vonGersdorff:2002us}. 

For the most part, bulk contributions to the axion potential on the interval are suppressed by $e^{-2 m \ell}$, so that the exponential suppression of quality-violating effects is the same as for the circle compactification. In our construction, contributions scaling as $e^{-m \ell}$ from Wilson lines linking source terms localized on the fixed planes \cite{Choi:2026kxu} are forbidden by the bulk $SU(2)$ and its remnants at the fixed planes. Generating these source terms would require adding bulk matter fields neutral under the $\Z_2$ center of $SU(2)$ with components that survive projection on both fixed planes.

\subsection{Monopole string parallel to the interval}
\label{sec:orbifold-parallel}

A string parallel to the interval would again have worldsheet $(t,y)$, which gave a stable four-dimensional monopole on the circle. On the orbifold, however, the absence of a massless $U(1)$ zero mode presages a problem:  there is
no analogous smooth finite-energy charged particle. 

To see this, consider a large cylinder $S_R^2\times[0,\ell]$ surrounding a putative interval-crossing segment. For each $y$ define the monopole degree
\begin{equation}
  m(y)=\frac{1}{2\pi}\int_{S_R^2\times\{y\}}\cF.
  \label{eq:monopole-degree-y}
\end{equation}
Away from zeros of $\Phi$, the gauge-invariant abelian field strength obeys $\dd\cF=0$,
so $m(y)$ is independent of $y$. But on a fixed plane, the allowed Higgs directions lie in the $T^2$-$T^3$ equator and the allowed boundary
gauge field is along $T^1$. Thus the restriction of the 't Hooft electromagnetic two-form (see Appendix \ref{app:setup}) to the fixed plane vanishes: the $\hat\Phi^a F^a$ term has no component along
$\hat\Phi$, and the target-space area form
$\epsilon^{abc}\hat\Phi^aD\hat\Phi^b\wedge D\hat\Phi^c$ vanishes when $\hat\Phi$ is
restricted to the equator. Hence
\begin{equation}
  m(0)=m(\ell)=0,
\end{equation}
and therefore $m(y)=0$ for all $y$. A net unit monopole charge cannot be carried by an
isolated interval-crossing string segment. Projecting out
$\cA_\mu^{(0)}$ therefore does not leave behind a stable monopole in the conventional sense.

\subsection{Monopole string perpendicular to the interval}
\label{sec:orbifold-perpendicular}

For the orbifold, an axion string arises when a monopole string lies along a noncompact direction $z$ parallel to the
fixed planes. The transverse space is
\begin{equation}
  M_\perp=\R^2_{\rho,\varphi}\times I_y,
  \qquad 0\leq y\leq \ell,
\end{equation}
while the linking surface is a cylinder $S^1_\varphi\times I_y$ rather than a torus. Because the Cartan component $A_\varphi^3$ (equivalently $\cA_\varphi$) is odd under the orbifold, it vanishes at the fixed planes. For the orbifold axion
$\vartheta=2\int_0^\ell\cA_y\dd y$,
\begin{align}
  \oint_{S^1_\varphi}\dd\vartheta
  &=2\oint\dd\varphi\,\partial_\varphi\int_0^\ell\dd y\,\cA_y
  =2\int_{S^1_\varphi\times I_y}\cF.
\end{align}
The integer winding is therefore the twisted first Chern class
\begin{equation}
  w(\gamma)
  =\frac{1}{2\pi}\oint_\gamma\dd\vartheta
  =\frac{1}{\pi}\int_{\gamma\times I_y}\cF
  =\frac{1}{2\pi}\int_{\gamma\times S^1_{\rm cover}}\cF
  \in\Z.
  \label{eq:orbifold-winding}
\end{equation}
A convenient large-distance representative for a winding-$w$ string is
\begin{equation}
  \cA_y\simeq \frac{w}{2\ell}\varphi,
  \qquad
  \cA_\varphi\simeq0,
  \qquad
  \vartheta(\varphi)=w\varphi,
\end{equation}
understood patchwise. Then
\begin{equation}
  \cF_{\varphi y}\simeq \frac{w}{2\ell},
  \qquad
  B_\rho\simeq \frac{w}{2\ell\rho}.
\end{equation}
The tail energy is
\begin{equation}
  \mu_{\rm tail}^{\rm orb}
  =\pi w^2 f_{\rm orb}^2\log\frac{\rho_{\rm IR}}{\rho_{\rm UV}},
  \qquad
  f_{\rm orb}^2=\frac{1}{4g_{5,A}^2\ell}.
  \label{eq:orbifold-tail-tension}
\end{equation}
Thus the orbifold perpendicular monopole string is again a four-dimensional global
axion string, but the winding and core differ from the naive
abelian-interval expectation.

The most useful way to see the defect spectrum is to use the covering circle of length
$2\ell$. An orbifold solution is a $\Z_2$-symmetric periodic configuration. Since
$r^*\cA=-\cA$ and $r^*\cF=-\cF$ under the reflection $r:y\mapsto -y$, a bulk monopole
string parallel to the fixed planes has a same-charge mirror on the cover: the reflection
flips both the field strength and the orientation of the small linking two-sphere. Therefore
a smooth full monopole string in the interior of the interval carries
\begin{equation}
  w=\pm2.
\end{equation}
The fundamental interval contains one physical bulk core, but in the orbifold that
core is a winding-two axion string.

The primitive $w=\pm1$ string is instead a fixed-plane half-monopole. This object is the smooth quotient of an ordinary
't Hooft--Polyakov monopole string on the doubled cover, centered on the fixed plane and
oriented consistently with the orbifold parity. For a string along $z$ at $y=0$, take the
ordinary regular monopole ansatz of Sec.~\ref{sec:noncompact-monostring} with transverse
coordinates
\begin{equation}
  X^1=y,\qquad X^2=x,\qquad X^3=x_\perp,
\end{equation}
and identify the internal indices with the same labels. Then $\Phi^1\propto y$ is odd,
while $\Phi^{2,3}$ are even, as required by \eqref{eq:scalar-parity}. Likewise the gauge
field components obey the parities in \eqref{eq:gauge-parity}. The quotient $y\geq0$ is a
smooth half-space soliton. An analogous construction gives a half-monopole at the
$y=\ell$ fixed plane. The fact that minimal-winding strings live at the fixed planes makes it apparent why the fixed planes must contribute to the inflow that cancels the anomalies of fermion zero modes living on the strings.

The analytic obstruction to a closed-form compact-space solution is the same as in the
circle perpendicular case, with the additional orbifold boundary conditions. The full
Yang-Mills-Higgs equations are nonlinear elliptic PDEs on
$\R^2_{\rho,\varphi}\times I_y$. However, the asymptotic construction is robust in the
regime $m_W\ell,m_\rho\ell\gg1$: the nonabelian core is locally the appropriate full or
half 't Hooft--Polyakov string, the intermediate abelian region is controlled by magnetic
flux, and the far field is the global string of the zero-mode axion $\vartheta$.

For $m_W\ell,m_\rho\ell\gg1$, the unit fixed-plane string has tension
\begin{equation}
  \mu_{w=1}^{\rm fp}(\rho_{\rm IR})
  =
  \frac{T_M}{2}+\Delta\mu_{\rm fp}
  +\pi f_{\rm orb}^2\log\frac{\rho_{\rm IR}}{\ell}
  +\cdots,
  \label{eq:fixed-plane-string-tension}
\end{equation}
with $\Delta\mu_{\rm fp}\sim \order(1/(g_{5,A}^2\ell))$. A smooth interior bulk core has
\begin{equation}
  \mu_{w=2}^{\rm bulk}(\rho_{\rm IR})
  =
  T_M+\Delta\mu_{\rm bulk}
  +4\pi f_{\rm orb}^2\log\frac{\rho_{\rm IR}}{\ell}
  +\cdots.
  \label{eq:bulk-w-two-tension}
\end{equation}
Because the long-distance energy scales as $w^2$, a winding-two bulk string is not
expected to be stable when splitting into unit-winding strings is dynamically allowed. It can
lower its global-gradient energy by fissioning into two unit fixed-plane strings, possibly on
the same fixed plane or on opposite fixed planes, while preserving the total winding.

As for the unit strings, these are pinned to the boundaries in the ultraviolet but are ordinary global axion
strings in the infrared. They cannot simply move off the fixed plane as an isolated smooth
bulk monopole, since moving a unit core into the interior produces a same-charge mirror and doubles the winding. In principle, they may switch from one fixed plane to the other through a finite-energy interval-crossing segment attached to the string, i.e. a bead or kink. In any event, unit orbifold axion strings are stable as
four-dimensional EFT strings until the $SU(N)$ instanton potential generates domain walls. These are the strings we're looking for.

\section{QCD axion string cosmology}
\label{sec:string-cosmology}

The cosmology of these strings is close to, but not identical to, ordinary post-inflationary
axion-string cosmology. At distances larger than the compactification scale the object
is a global string of the Wilson-line axion, while its ultraviolet core is a compactified five-
dimensional monopole string.
In this section we specialize the color group to QCD, $SU(3)_C$. We focus on the orbifold compactification because it removes the additional light fields and
4d magnetic monopole states of the circle spectrum while retaining the monopole-core axion
string. In this section we denote the
low-energy axion decay constant by $f_a$, equal to $f_{\rm orb}$ in
Eq.~\eqref{eq:forb}.

\subsection{Regime of validity}

We begin by specifying the hierarchy of scales that defines the cosmological
regime of interest.  We assume from the outset that the orbifold interval has
already been stabilized at fixed proper length $\ell$.  The fixed-interval
approximation is valid when the radion is heavy compared with the cosmological
time scales, so that its finite-temperature displacement is negligible.  The
details of the radion EFT and stability conditions are collected in
Appendix~\ref{app:radion-eft}.  Throughout this section $\ell$ denotes the
stabilized proper size of the extra dimension. 

We assume the inflationary epoch itself is four-dimensional, so that the
inflationary Hubble scale $H_I$ lies below the interval compactification scale,
\begin{equation}
  H_I<\ell^{-1}.
  \label{eq:four-dimensional-inflation-bound}
\end{equation}
Inflation itself could also be five-dimensional, but we leave this scenario to
future work.  After inflation, the gravitational background varies
on the Hubble scale
and never exceeds the Hubble scale at the end of inflation, $H_{\rm end}$.
The regime of interest is therefore hybrid: the long-wavelength gravitational
background does not resolve the compact direction,
\begin{equation}
  H\leq H_{\rm end}\ll m_{\rm KK}\sim\ell^{-1},
  \label{eq:hybrid-cosmology-regime}
\end{equation}
where $m_{\rm KK}$ is the Kaluza-Klein scale of the interval.  By contrast, the
thermal sector is controlled by the post-inflationary bath, whose temperature
peaks at $T_{\max}^{5d}$ and later reaches the symmetry-restoring phase transition at
$T_*$.  We require this bath to resolve the compact direction throughout the
transition while remaining below the five-dimensional EFT cutoff $\Lambda_5$,
\begin{equation}
  \ell^{-1}<T_*\leq T_{\max}^{5d}<\Lambda_5.
  \label{eq:thermal-compact-resolution-condition}
\end{equation}

Equation~\eqref{eq:hybrid-cosmology-regime} is the condition that permits a
four-dimensional treatment of gravity: external derivatives on the homogeneous
metric are of order $H$, while the nonzero gravitational KK modes have masses
of order $\ell^{-1}$.  Their effects on the background are therefore
suppressed by powers of $H\ell$, as in related stabilized-radion reductions of
higher-dimensional gravity~\cite{Giudice:2002vh}.

The same compactified KK description also explains why the thermal bath scales
five-dimensionally when $T\ell\gg1$. 
The energy density per ordinary three-volume is
\begin{equation}
  \rho_{4d}
  =
  N_{\rm th}^{(5)}\sum_n\int\frac{\dd^3p}{(2\pi)^3}
  \frac{E_{n,p}}{e^{E_{n,p}/T}\mp1},
  \qquad
  E_{n,p}=\sqrt{\mathbf p^2+m_n^2},
  \label{eq:thermal-kk-energy-density}
\end{equation}
where $N_{\rm th}^{(5)}$ denotes the effective number
of five-dimensional thermal degrees of freedom, and with the sign chosen according to statistics.  If $T\ll\ell^{-1}$, only the
zero mode is populated and the usual four-dimensional scaling
$\rho_{4d}\propto T^4$ is recovered.  If instead $T\ell\gg1$, many KK levels
are populated and the sum over modes is well approximated by a continuum
integral,
\begin{equation}
  \sum_n\longrightarrow
  \ell\int\frac{\dd p_y}{2\pi},
  \label{eq:thermal-kk-continuum}
\end{equation}
up to order-one orbifold and boundary-condition factors that we absorb
into the effective number of degrees of freedom.  Hence
\begin{equation}
  \rho_{4d}
  \simeq
  N_{\rm th}^{(5)}\,\ell
  \int\frac{\dd^3p\,\dd p_y}{(2\pi)^4}
  \frac{\sqrt{\mathbf p^2+p_y^2}}
  {e^{\sqrt{\mathbf p^2+p_y^2}/T}\mp1}
  \equiv
  \ell\,\rho_{5d} = \ell\,   \frac{3\zeta(5)}{\pi^2}\,N_{\rm th}^{(5)}T^5
  \label{eq:thermal-five-dimensional-phase-space}
\end{equation}
where $\zeta$ is the Riemann zeta function.  If gravity-sector KK particles are also
thermally populated, their contribution can be included in $N_{\rm th}^{(5)}$; in
weakly coupled gravity their abundance is model-dependent and need not be
thermal.   
Using the four-dimensional Friedmann equation with the reduced Planck mass,
\begin{equation}
  H^2 = \frac{\rho_{4d}}{3M_{\rm Pl}^2},
\end{equation}
therefore gives
\begin{equation}
  H^2(T)
  \simeq
  \frac{\zeta(5)N_{\rm th}^{(5)}\ell T^5}{\pi^2 M_{\rm Pl}^2}.
\end{equation}
This gives a useful estimate of the maximum temperature: for optimistic instantaneous reheating,
with $H_{\rm end}$ taken to be comparable to $H_I$ in the estimates below,
\begin{equation}
  T_{\max}^{5d}
  =
  \left[
    \frac{\pi^2}{\zeta(5)N_{\rm th}^{(5)}}
    M_{\rm Pl}^2H_{\rm end}^2\ell^{-1}
  \right]^{1/5}.
  \label{eq:maximum-five-dimensional-temperature}
\end{equation}  The necessary condition that
the maximum temperature resolve the interval, $T_{\max}^{5d}>\ell^{-1}$, gives
\begin{equation}
  \ell^{-1}
  \lesssim
  \left[
    \frac{\pi^2}{\zeta(5)N_{\rm th}^{(5)}}
  \right]^{1/4}
  \left(M_{\rm Pl}H_{\rm end}\right)^{1/2}.
\end{equation}
If $H_{\rm end}\simeq H_I$ saturates the CMB tensor limit $H_I/M_{\rm Pl}\lesssim 10^{-5}$~\cite{BICEPKeck:2021gln}, this compactification-resolution bound permits compactification scales much higher than the range we will consider for QCD axion dark matter below.  For lower-scale inflation, however, the same inequality can become the relevant upper bound on $\ell^{-1}$ and should be evaluated with the corresponding $H_{\rm end}$.

\subsection{String network formation and scaling}

The hierarchy above is necessary but not sufficient for producing the desired
defects. Denoting the correlation length at formation
by $\xi_*$, and the Hubble scale at formation by $H_*$, we
focus on the locally five-dimensional formation regime
\begin{equation}
  \xi_*<\ell<H_*^{-1}.
  \label{eq:local-five-dimensional-formation-window}
\end{equation}
For the monotonic histories considered here, the right-hand inequality follows
from $H_*\leq H_{\rm end}$ and
Eq.~\eqref{eq:hybrid-cosmology-regime}.  The nontrivial microscopic
requirement is $\xi_*<\ell$: correlated domains must be smaller than the
interval in order to form topologically nontrivial gauge configurations across
the compact direction.  A continuous transition can satisfy this if the
freezeout length $\xi_{\rm KZ}$ is set by Kibble-Zurek dynamics rather than the
horizon. 
In terms of microscopic correlation-length and relaxation-time
scales $\xi_0$ and $\tau_0$, quench time $\tau_Q$, critical exponents $\nu$
and $z$, and characteristic microscopic mass $m_*$,
\begin{equation}
  \xi_{\rm KZ}
  \sim
  \xi_0
  \left(
    \frac{\tau_Q}{\tau_0}
  \right)^{\frac{\nu}{1+z\nu}}.
  \label{eq:kibble-zurek-correlation-length}
\end{equation}
 For a first-order transition, the analogous length is
set by the bubble nucleation and collision scale rather than by the horizon; for
rapid transitions this can also give $\xi_*\ll H_*^{-1}$~\cite{Linde:1981zj}.
The estimates below assume that the microscopic transition supplies such a
short freezeout length.

The phase transition first produces smooth bulk monopole
strings.  On the orbifold these carry Wilson-line winding $w=2$.  The
long-lived cosmological network, however, should be built from the primitive
unit strings.  In the smooth minimal orbifold sector there is no light
zero-mode endpoint that would let unit-winding fixed-plane strings break after
they form.  We therefore assume that the bulk strings fission into unit-winding
$w=1$ fixed-plane strings before the QCD axion potential turns on.  A simple
classical estimate suggests that this is a microscopic process in the
controlled regime.  Separating the two unit strings by a distance of order
$\ell$ gives a repulsive force per unit length
$F/L\sim f_a^2/\ell$.  With mass per unit length of order $T_M$, the
acceleration is
\begin{equation}
  a_{\rm fiss}
  \sim
  \frac{f_a^2}{T_M\ell}
  \sim
  \frac{1}{m_W\ell^2},
  \qquad
  t_{\rm fiss}\sim \ell\sqrt{m_W\ell},
  \label{eq:fission-timescale}
\end{equation}
up to order-one factors and details of the fixed-plane dynamics.  This estimate
uses the long-range $w^2$ gradient repulsion at separations larger than the
core size and assumes that the $w=2$ bulk core has no additional barrier to
splitting and migration to the fixed planes; away from the BPS point,
$m_\rho\neq m_W$, the short-distance repulsion of like-charge cores at distances
of order $m_\rho^{-1}$ supports this picture.  For
$m_W\ell$ in the semiclassical window considered below, this time is far
shorter than the Hubble time after formation.  We therefore treat the
long-lived network as a network of unit-winding fixed-plane strings.

After the transition, the thermal bath cools below the compactification
scale, $T\sim\ell^{-1}$, well before QCD in the parameter ranges considered
below, so thermally populated KK modes have decoupled by the time the axion
potential becomes important.  Together with $\ell<H_*^{-1}$, this means that
on distances larger than $\ell$ the fixed-plane strings are seen as ordinary
line defects in the noncompact three dimensions.  We model the long-distance
network by the global-string scaling form
\begin{equation}
  \mathcal L_4(t)\simeq \frac{\mathcal S(t)}{t^2},
  \qquad
  \rho_{\rm string}(t)\simeq \mu(t)\mathcal L_4(t).
  \label{eq:global-string-scaling}
\end{equation}
where $\mathcal{L}_4$ is the string length density and $\mathcal S(t)$ measures the total string length in a horizon-scale
volume, up to order-one factors relating $t$ and $H^{-1}$.  For a unit-winding
fixed-plane string, Eq.~\eqref{eq:fixed-plane-string-tension} with infrared
radial cutoff $\rho_{\rm IR}\sim t$ can be packaged as an effective logarithm,
\begin{equation}
  \kappa_{\rm eff}^{\rm orb}(t)
  \equiv \frac{\mu_{\rm orb,unit}(t)}{\pi f_a^2}
  \simeq
  2m_W\ell\,B(m_\rho/m_W)+\log\frac{t}{\ell}+\cdots .
  \label{eq:orbifold-effective-logarithm}
\end{equation}
Here the first term is the half-monopole core tension in units of
$\pi f_a^2$, using $f_a^2=1/(g_{5,2}^2\ell)$ and
Eq.~\eqref{eq:monostring-tension}.  The quality and cutoff requirements
discussed above make this core contribution large enough to affect the network,
but not arbitrarily large.  Existing axion-string simulations find
$\mathcal S=O(1)$ values.  In the present case, however, the UV physics of the monopole core changes could potentially affect the dynamics of string reconnection and loop production as well as decay of heavy modes from the core beyond the predictions of existing simulations. In particular, as discussed in Secs.~\ref{sec:fermion-cs}
and~\ref{sec:orbifold-perpendicular}, the monopole-string core carries an
anomalous chiral zero mode charged under $SU(3)_C$.  Such colored worldsheet
modes can in principle make the strings current-carrying, affecting current
buildup, intercommutation, loop evolution, and possible vorton-like loop
stabilization.  We do not attempt a transport calculation for these modes here.
Our expectation is that, in the minimal setup, they are not associated with a
long-lived conserved current after confinement: the same QCD dynamics that
generates the axion potential and attaches walls also gaps or confines the
colored zero modes. 

\subsection{Dark matter production}
We estimate the relic density from the axion number produced by the scaling
string network before the QCD epoch, following the formalism of Appendix F of
Ref.~\cite{Buschmann:2021sdq}.  The relevant matching temperature, denoted here
by $T_a$, is defined by
\begin{equation}
  3H(T_a)=m_a(T_a),
  \qquad H_a\equiv H(T_a).
  \label{eq:qcd-string-matching-temperature}
\end{equation}
$T_a$ is the temperature at which the QCD
axion mass becomes dynamically important for the string-wall network (not
merely the homogeneous-misalignment onset temperature).  The assumption is that the network evolves in scaling until $T_a$, rapidly
collapses or evaporates afterwards, and leaves behind a conserved axion number.

We determine the contribution to the axion number density from the string network at $T_a$ in terms of the network density,
tension, and emission spectrum:
\begin{equation}
  n_a^{\rm str}(T_a)
  \simeq
  \frac{8\pi f_a^2 H_a}{\delta_a}
  \sqrt{\mathcal S_a}\,\kappa_a,
  \qquad
  \kappa_a\equiv\frac{\mu_a}{\pi f_a^2},
  \label{eq:appendix-f-string-number}
\end{equation}
where $\mathcal S_a\equiv\mathcal S(T_a)$ is the string length per Hubble
volume, corresponding to the $\xi_*$ convention of
Refs.~\cite{Buschmann:2021sdq,Benabou:2024amr}, and $\kappa_a$ is the
dimensionless string tension evaluated at the horizon scale $H_a^{-1}$.  For an
ordinary global string, $\kappa_a$ is the familiar large logarithm $\log \left( m_r / H_a \right)$, where $r$ is the mass of the radial mode for the PQ axion.  The spectral parameter $\delta_a$ is fixed by
the normalized distribution for physical momentum $k$,
\begin{equation}
  F(k/H)
  \equiv
  \frac{d\log\Gamma_a}{d(k/H)},
\end{equation}
where $\Gamma_a$ is the axion emission rate from the scaling network.  Averaging
over this distribution gives
\begin{equation}
  \left\langle\frac{H}{k}\right\rangle_F^{-1}
  =
  \delta_a\sqrt{\mathcal S_a}.
\end{equation}
For the $F(k/H)\propto(k/H)^{-q}$ fit with $q=1.02$ in
Ref.~\cite{Benabou:2024amr}, $\delta_a\simeq318$.  

The value of $q$ is also where the adaptive mesh refinement (AMR) simulations \cite{Benabou:2024amr, Buschmann:2021sdq} differ most significantly from the
static-lattice extrapolation of Refs.~\cite{Gorghetto:2018myk,Gorghetto:2020qws}.
Both analyses find a growing string density, but they differ in the inferred
late-time behavior of the axion emission spectrum.  Refs.~\cite{Gorghetto:2018myk,Gorghetto:2020qws}
find evidence that the fitted spectral index grows logarithmically with
$m_{r}/H$, leading to a more infrared-dominated extrapolation and a
larger string contribution to the relic abundance.  By contrast,
Ref.~\cite{Benabou:2024amr} finds no statistically significant logarithmic
growth and is consistent with an approximately scale-invariant emission
spectrum.  In the rescaling estimates below we use the AMR simulation values as
fiducial parameter points, without attempting to determine which extrapolation
of the emission spectrum is correct. 

The final abundance comes from redshifting this axion number to the present epoch.
With the simulation parameters from Ref.~\cite{Benabou:2024amr}, the
string/network contribution for the PQ axion strings is:
\begin{equation}
  \Omega_a^{\rm str}h^2
  \simeq
  0.12
  \left(
    \frac{f_a}{1.4\times10^{11}\,{\rm GeV}}
  \right)^{1.17}
  \left(
    \frac{318}{\delta_a}
  \right)
  \sqrt{\frac{\mathcal S_a}{13}}
  \left(
    \frac{\kappa_a}{70}
  \right).
  \label{eq:amr-string-abundance}
\end{equation}

We now estimate the corresponding scaling for the orbifold strings, from which we expect the dominant change to be from the new effective string tension:
\begin{equation}
  \kappa_a^{\rm orb}
  =
  \frac{\mu_{\rm orb,unit}(H_a^{-1})}{\pi f_a^2}
  \simeq
  2m_W\ell\,B(m_\rho/m_W)
  +\log\frac{H_a^{-1}}{\ell}
  +\cdots,
  \label{eq:orbifold-kappa-at-qcd}
\end{equation}
where we have used Eq.~\eqref{eq:orbifold-effective-logarithm} with the infrared cutoff set
by the horizon at $T_a$.  This is the key quantity to compare with the ordinary
global-string value used in the simulations; at fixed $\delta_a$ and
$\mathcal S_a$, the AMR number density is proportional to $\kappa_a$.

We match to an ordinary
four-dimensional PQ axion string using the same infrared cutoff.  With
$\kappa_{\rm std}\equiv\log(H_a^{-1}m_{\rm UV}^{\rm std})\simeq70$ for a
standard UV scale $m_{\rm UV}^{\rm std}$,
\begin{equation}
  \kappa_a^{\rm orb}
  \simeq
  2m_W\ell\, B(m_\rho/m_W)
  +\kappa_{\rm std}
  -\log(m_{\rm UV}^{\rm std}\ell)
  +\cdots .
\end{equation}
The subtraction is a logarithmic matching ambiguity associated with the onset of
the four-dimensional tail at $\ell$; the parametric effect is the monopole-core
term proportional to $m_W\ell$.  For $B\simeq1$ and
$10\lesssim m_W\ell\lesssim10^3$,
\begin{equation}
  \frac{\kappa_a^{\rm orb}}{70}
  \sim
  1\text{--}30,
\end{equation}
If $\delta_a^{\rm orb}$ and $\mathcal S_a^{\rm orb}$ remain close to the AMR
reference values, the network-scaling contribution suggests
\begin{equation}
  m_{a,0}^{\rm orb}
  \sim
  10^{-4}\text{--}10^{-3}\,{\rm eV},
  \label{eq:orbifold-amr-estimate}
\end{equation}
up to the network-level uncertainties in $\mathcal S_a^{\rm orb}$,
$\delta_a^{\rm orb}$, and radiation into non-axion degrees of freedom.

The complete abundance also contains terms not included in this network-scaling
estimate,
\begin{equation}
  \Omega_a^{\rm full}
  =
  \Omega_a^{\rm str}
  +\Omega_a^{\rm mis}
  +\Omega_a^{\rm coll}
  +\cdots .
\end{equation}
Here $\Omega_a^{\rm str}$ denotes the network-scaling string contribution
estimated above, $\Omega_a^{\rm mis}$ is the homogeneous misalignment
contribution from coherent oscillations of the axion field, and
$\Omega_a^{\rm coll}$ denotes axions radiated during the final QCD-era collapse
of the string-wall network.  For the post-inflationary scenario, the standard
average over initial misalignment angles gives~\cite{Borsanyi:2016ksw,Bae:2008ue,Wantz:2009it,Ballesteros:2016euj}
\begin{equation}
  \Omega_a^{\rm mis}h^2
  \simeq
  0.03
  \left(
    \frac{0.1\,{\rm meV}}{m_{a,0}}
  \right)^{1.165}.
  \label{eq:standard-misalignment-abundance}
\end{equation}
It was suggested in  \cite{Buschmann:2021sdq} that once $m_a(T)\gg H$, the axion
mass provides an infrared cutoff that pushes radiation from string-wall network collapse toward higher
physical momenta than the scaling-network spectrum. At fixed energy density, this
produces fewer axions by number, so we leave $\Omega_a^{\rm coll}$ as an
uncertain additional contribution rather than assigning it a numerical
rescaling.

The mass range of Eq.~\ref{eq:orbifold-amr-estimate} approximately fixes the product of the extra-dimensional size and
five-dimensional gauge coupling.  From Eq.~\eqref{eq:forb}, the Cartan
normalization in Appendix~\ref{app:setup} and the relation
$m_{a,0}^2f_a^2=\chi_{\rm QCD}$,
\begin{equation}
  g_{5,2}^2\ell
  =
  \frac{1}{f_a^2}
  \simeq
  \frac{m_{a,0}^2}{\chi_{\rm QCD}},
  \qquad
  10^{10}\,{\rm GeV}
  \lesssim
  f_a
  \lesssim
  10^{11}\,{\rm GeV}.
\end{equation}
Here $\chi_{\rm QCD}$ is the zero-temperature QCD topological susceptibility.
To express this as a compactification scale, we trade the five-dimensional
coupling for the corresponding four-dimensional monopole-sector coupling
evaluated at the compactification scale,
\begin{equation}
  g_{4,2}^2(\ell^{-1})\equiv \frac{g_{5,2}^2}{\ell}.
\end{equation}
Keeping this matching coupling below the QCD coupling at the compactification scale
prevents the monopole sector from further lowering the five-dimensional NDA
cutoff in Eq.~\eqref{eq:cutoff}:
\begin{equation}
  g_{4,2}(\ell^{-1})\lesssim g_s(\ell^{-1}),
  \qquad
  g_s^2(\ell^{-1})\equiv4\pi\alpha_s(\ell^{-1}) .
\end{equation}
Using one-loop QCD running with $\alpha_s(1\,{\rm TeV})\simeq0.09$ and solving
$x=g_s(x)f_a$ for $x\equiv\ell^{-1}$ over the AMR range gives the
order-of-magnitude upper bound
\begin{equation}
  \ell^{-1}
  \lesssim
  g_s(\ell^{-1})\,f_a
  \sim
  10^{10}\text{--}10^{11}\,{\rm GeV}.
  \label{eq:compactification-amr-range}
\end{equation}

The same EFT requirement in Eq.~\eqref{eq:thermal-compact-resolution-condition} also prevents the compactification scale from being too low. This is consistent with the microscopic thermal restoration estimate: a five-dimensional
thermal mass,
$\Delta m_\Phi^2\sim g_{5,2}^2T^3$, gives
\begin{equation}
  T_*\ell
  \sim
  \left(\frac{m_W\ell}{g_{4,2}(\ell^{-1})}\right)^{2/3}.
  \label{eq:eta-thermal-restoration}
\end{equation}
Since $g_{4,2}(\ell^{-1})=\ell^{-1}/f_a$, lowering $\ell^{-1}$ weakens the monopole-sector coupling and raises the transition temperature in compactification units.  Meanwhile, with QCD setting the NDA cutoff in Eq.~\eqref{eq:cutoff}, $\Lambda_5\simeq[8\pi^3/g_s^2(\ell^{-1})]\ell^{-1}$, where the prefactor varies only logarithmically.  Requiring $T_*\lesssim\Lambda_5$ therefore gives
\begin{equation}
  \ell^{-1}
  \gtrsim
  f_a\,(m_W\ell)
  \left[\frac{g_s^2(\ell^{-1})}{8\pi^3}\right]^{3/2}.
  \label{eq:compactification-lower-bound}
\end{equation}
Assuming $m_W$ is the relevant scale for the exponentially-suppressed quality violation,
\begin{equation}
  \ell_{\rm min}^{-1}
  \sim
  5\times10^7\,{\rm GeV}
  \left(\frac{f_a}{10^{10}\,{\rm GeV}}\right)
  \left(\frac{m_W\ell}{50}\right)
  \left(\frac{g_s^2(\ell^{-1})}{0.5}\right)^{3/2},
\end{equation}
so the axion dark matter range above suggests a rough lower edge
$\ell^{-1}\gtrsim10^8$--$10^9\,{\rm GeV}$.

The reheating check uses the stronger transition condition in
Eq.~\eqref{eq:thermal-compact-resolution-condition}.
For $10\lesssim m_W\ell\lesssim10^3$ and
$g_{4,2}(\ell^{-1})\lesssim g_s(\ell^{-1})\lesssim1$, Eq.~\eqref{eq:eta-thermal-restoration} naturally gives $T_*\ell$
of order $10$--$100$ over much of the controlled range.
Taking $H_{\rm end}\simeq H_I$, the
reheating lower bound on $H_{\rm end}$ may also be read as a lower bound on the
inflationary scale.  For $N_{\rm th}^{(5)}\sim100$, evaluating this condition
over the compactification range in Eq.~\eqref{eq:compactification-amr-range}
gives
\begin{equation}
  H_{\rm end}
  \gtrsim
  10^{4}\text{--}10^{6}\,{\rm GeV}
  \left(\frac{T_*\ell}{10}\right)^{5/2}.
\end{equation}
Combining this with the fixed-size hierarchy $H_{\rm end}\ll\ell^{-1}$, the
order-of-magnitude reheating window for $T_*\ell=10$--$100$ is
\begin{equation}
  10^{4}\text{--}10^{9}\,{\rm GeV}
  \lesssim
  H_{\rm end}
  \ll
  10^{10}\text{--}10^{11}\,{\rm GeV}.
\end{equation}
The corresponding transition temperatures are
$T_*\sim10^{11}$--$10^{13}\,{\rm GeV}$.

This suggestive axion dark matter mass range is useful to compare with pre-inflationary
misalignment.  For an order-one initial angle, the usual misalignment estimate
points to QCD axion masses around the $\mu{\rm eV}$ scale, but could potentially select much lower masses for small initial misalignment angles.  Conversely, tuning the initial
angle close to the top of the potential can raise the dark matter mass for pre-inflationary misalignment
toward the meV scale, but this regime is increasingly constrained by
isocurvature \cite{Huang:2020etx, Arvanitaki_2020, Planck:2018jri}.  On the high-mass side, stellar-cooling and supernova bounds
constrain QCD axions roughly from the $10^{-2}\,{\rm eV}$ masses upward, with
model-dependent variations.  The extra-dimensional string histories above
therefore help populate the intermediate mass range where a viable dark matter
production mechanism is otherwise more challenging~\cite{Preskill:1982cy,Abbott:1982af,Dine:1982ah,DiLuzio:2020wdo}.

\section{Conclusions}
\label{sec:conclusion}

In this paper, we have established a simple field-theoretic interpretation of extra-dimensional axion strings: they can arise from five-dimensional 't Hooft--Polyakov monopole strings. The relevant phase transition involves spontaneous breaking of a 5d $SU(2)$ gauge symmetry rather than a 4d PQ global symmetry. The fifth component of the unbroken $U(1)$ gauge symmetry becomes the 4d axion, while the monopole strings perpendicular to the extra dimension become axion strings. The resulting axion still enjoys the high quality of an extra-dimensional axion, albeit with irreducible (but tolerable) contributions to its potential arising from the fact that its protective one-form global symmetry is emergent below the scale of $SU(2)$ breaking.

The detailed solitonic spectrum of the axion EFT depends on the compactification. Under circle compactification, monopole strings parallel to the extra dimension become monopoles and monopole rings under the surviving abelian gauge theory, while bulk monopole strings perpendicular to the extra dimension become minimal-winding axion strings. In contrast, compactification on an orbifold preserves the axion but eliminates the corresponding 4d vector and magnetically charged solitons. Monopole strings perpendicular to the extra dimension survive to become axion strings, with minimally-wound strings localized on the fixed planes and doubly-wound strings living in the bulk. 

As the relevant scales of the strings are associated with the five-dimensional gauge theory, rather than gravity, this allows for greater parametric scale separation compared to fundamental strings. In a simple cosmology where inflation is effectively four-dimensional but reheating accesses the fifth dimension and restores the bulk $SU(2)$ symmetry, monopole strings are produced as the universe cools. Their subsequent evolution is close to the familiar 4d post-inflationary axion string cosmology and produces a viable axion dark matter abundance. 
This blurs the historical dichotomy between axion models with extra-dimensional solutions to the quality problem and those with predictive post-inflationary axion string cosmology.

There are many open directions for further investigation. For the sake of clarity we have primarily focused on UV models leading to the axion--Yang Mills effective theory at low energies, but it would be fruitful to construct fully realistic models for the QCD axion. Another important aspect of realism is the 5d spacetime: the modest hierarchy between the compactification scale and UV cutoff in a flat extra dimension suggests exploring warped compactifications \cite{CraigMishraToAppear}. There are many open questions associated with the early-universe cosmology of these constructions, including the detailed cosmological string evolution of our four-dimensional inflationary scenario and the parametrics of effectively five-dimensional scenarios.

The potential phenomenological relevance of five-dimensional monopole strings also motivates further study. We have argued for the existence and properties of our axion strings starting from exact solutions in five extended dimensions and parametric arguments under compactification, but full numerical solutions for the solitons would be invaluable. While our primary focus has been on the monopole strings perpendicular to the compact dimension, the phenomenology of the magnetically charged excitations arising from strings wrapping the circle is potentially rich. More broadly, despite decades of work on phenomenological models of extra dimensions, the properties and fates of extra-dimensional solitons in these models remain largely untouched and ripe for exploration.

\section*{Note added}

Throughout the course of this research we have been aware of parallel and independent work on the same topic by Rudin Petrossian-Byrne and Giovanni Villadoro, to appear simultaneously with this manuscript.

\acknowledgments

We thank Prateek Agrawal, Joshua Benabou, Isabel Garcia Garcia, Elliott Gesteau, Alexandre Homrich, Marius Kongsore, John March-Russell, Matt Reece, Mario Reig, Ben Safdi, Dan Sehayek, and Mykhaylo Usatyuk for useful conversations, and Prateek Agrawal \& Dan Sehayek for comments on the manuscript. We particularly thank Rudin Petrossian-Byrne and Giovanni Villadoro for both useful conversations and coordination of our related work. The work of NC was supported in part by the U.S. Department of Energy under the grant DE-SC0011702. The work of AM was supported by grant GBMF7392 from the Gordon and Betty
Moore Foundation. This work was performed in part at the Kavli Institute for Theoretical Physics, supported by the National Science Foundation under the grant NSF PHY-1748958 and at the Aspen Center for Physics, which is supported by National Science Foundation grant PHY-2210452.

\appendix
\section{Setup and conventions \label{app:setup}}

We use mostly-minus Lorentzian signature.  For the bulk $SU(2)$ factor, we mainly work in terms of canonically normalized fields
\begin{equation}
  \cL\supset -\frac14 F^a_{MN}F^{aMN}+\frac12 D_M\Phi^aD^M\Phi^a,
  \qquad
  D_M\Phi^a=\partial_M\Phi^a+g_{5,2}\epsilon^{abc}A^b_M\Phi^c.
\end{equation}
The classical dimensions of fields and couplings are
\begin{equation}
  [A_M]=[\Phi]=\frac32,
  \qquad [g_{5,2}]=[y_5]=-\frac12,
  \qquad [\lambda_5]=-1 \,
\end{equation}
where $y_5$ is the Yukawa coupling and $\lambda_5$ the quartic coupling of the adjoint scalar. When $\Phi$ acquires a vev $v_5\equiv |\vev{\Phi}|$, this induces masses for the $W^\pm$ bosons of $SU(2)/U(1)$, the Dirac fermions, and the radial mode of $\Phi$ of order
\begin{equation}
  m_W\simeq g_{5,2}v_5,
  \qquad
  m_\psi\simeq y_5v_5,
  \qquad
  m_\rho\simeq \sqrt{\lambda_5}\,v_5,
  \label{eq:physical-masses}
\end{equation}
up to order-one factors.  The effective four-dimensional vev associated with a zero mode over a
compactification region of length $\mathcal L$ is
\begin{equation}
  v_4=\sqrt{\mathcal L}\,v_5,
  \qquad
  \mathcal L=\begin{cases}
    L, & S^1,\\
    \ell, & S^1/\Z_2.
  \end{cases}
\end{equation}
This naturally leads to four-dimensional couplings
\begin{equation}
  g_{4,2}^2=\frac{g_{5,2}^2}{\mathcal L},
  \qquad
  y_4=\frac{y_5}{\sqrt{\mathcal L}},
  \qquad
  \lambda_4=\frac{\lambda_5}{\mathcal L},
\end{equation}
such that $m_W=g_{5,2}v_5=g_{4,2}v_4$ and $m_\psi=y_5v_5=y_4v_4$.

For the unbroken $U(1)$ that ultimately furnishes the four-dimensional axion, it is convenient to work in terms of the unit-doublet-charge connection $\cA_M$. In the abelian regime this connection is related to the canonically normalized Cartan gauge field by
\begin{equation}
  \cA_M\equiv \frac{g_{5,2}}{2}A_M^3.
\end{equation}
Then a fundamental doublet couples as $D_M=\partial_M+i\cA_M\sigma^3$ in the Cartan background and has charges $q=\pm1$, while the massive $W^\pm$ gauge bosons have charges $\pm2$. The abelian kinetic term is
\begin{equation}
  \cL_{U(1)}=-\frac{1}{4g_{5,A}^2}\cF_{MN}\cF^{MN},
  \qquad
  \cF=\dd\cA,
  \qquad
  g_{5,A}=\frac{g_{5,2}}{2}.
\end{equation}
In terms of the UV theory, we can write the 't Hooft 2-form as
\begin{equation}
  \cF
  \equiv
  \frac{g_{5,2}}{2}\widehat{\Phi}^{\,a}F^a
  -\frac14\epsilon^{abc}\widehat{\Phi}^{\,a}
  D\widehat{\Phi}^{\,b}\wedge D\widehat{\Phi}^{\,c},
  \qquad
  \widehat{\Phi}^{\,a}\equiv\frac{\Phi^a}{|\Phi|}.
\end{equation}

With this convention the circle Wilson line $\theta=\int_0^L \dd y\,\cA_5$ has period $2\pi$ in the presence of unit-charge fields, and the circle magnetic flux is quantized as $\int\cF=2\pi n$. On the $SU(2)$ orbifold, the half-interval holonomy $\theta_I=\int_0^\ell\dd y\,\cA_5$ has period $\pi$ because center-valued gauge transformations at the fixed planes are allowed. The $2\pi$-periodic interval axion is $\vartheta=2\theta_I$, and its winding is measured by
\begin{equation}
  w=\frac{1}{2\pi}\oint\dd\vartheta
   =\frac{1}{\pi}\int_{S^1_\varphi\times I_y}\cF\in\Z.
\end{equation}

For the $SU(N)$ field strength $G$, our conventions are such that
\begin{equation}
  \frac{1}{8\pi^2}\int \tr_{\bf N}G\wedge G\in \Z,
\end{equation}
where $\tr_{\bf N}$ is the trace in the fundamental representation.

\section{Fermion boundary conditions}
\label{app:fermion-bcs}

In this appendix we spell out the boundary conditions for the bulk fermion $\Psi\sim({\bf N},{\bf 2})$ that generates the mixed Chern--Simons term.    In the Cartan basis of the $SU(2)$ factor, write
\begin{equation}
  \Psi=\begin{pmatrix}\psi_+\\ \psi_-\end{pmatrix},
\end{equation}
where the unbroken abelian connection is the unit-doublet-charge connection
$\mathcal A_M=(g_{5,2}/2)A^3_M$.  After the adjoint scalar acquires the vev
$\langle \Phi^a\rangle=v_5\delta^{a3},$
the two Cartan components have charges and signed Yukawa masses
\begin{equation}
  (q_+,m_+)=(+1,+m_\Psi),\qquad
  (q_-,m_-)=(-1,-m_\Psi),
  \qquad m_\Psi\equiv y_5v_5 .
  \label{eq:cartan-fermion-charges-masses}
\end{equation}
 If a constant $SU(2)$-invariant bare mass is present on the circle, we have instead
\begin{equation}
  m_\pm=M\pm y_5v_5 .
\end{equation}

\paragraph{Circle compactification.}
On the circle there is no orbifold projection.  The simplest boundary condition is periodic,
\begin{equation}
  \Psi(x,y+L)=\Psi(x,y),
  \label{eq:circle-periodic-fermion}
\end{equation}
but more generally we may have
\begin{equation}
  \Psi(x,y+L)=e^{2\pi i\alpha_S}\Psi(x,y),
  \qquad \alpha_S=0,\frac12,
  \label{eq:circle-spin-structure}
\end{equation}
where $\alpha_S=0$ and $\alpha_S=1/2$ correspond to periodic and antiperiodic conditions, respectively.  Both four-dimensional chiralities obey the same boundary condition, so the circle
compactification does not impose a chiral projection.  Each Kaluza--Klein level gives a four-dimensional Dirac fermion.

In a constant Wilson-line background
\begin{equation}
  \theta(x)=\int_0^Ldy\,\mathcal A_5(x,y),
\end{equation}
the KK momenta of the two Cartan components are
\begin{equation}
p_{5,n,\pm}=\frac{2\pi(n+\alpha_S)\pm\theta}{L},
  \qquad n\in\Z .
  \label{eq:circle-fermion-momenta}
\end{equation}
Equivalently, in a gauge where the constant background $\mathcal A_5$ is gauged away,
the Wilson line appears as a twist in the boundary condition,
\begin{equation}
  \psi_\pm(x,y+L)
  =e^{2\pi i\alpha_S}e^{\pm i\theta}\psi_\pm(x,y).
  \label{eq:circle-wilson-line-twist}
\end{equation}
The four-dimensional Dirac operator at KK level $n$ contains the complex mass operator
\begin{equation}
  m_\pm+i\gamma_5p_{5,n,\pm},
\end{equation}
while the corresponding physical mass-squared is
\begin{equation}
  M^2_{n,\pm}=m_\pm^2+p_{5,n,\pm}^2 .
\end{equation}
Because $\theta$ is the holonomy around a closed circle, the charge-one spectrum is
invariant under
\begin{equation}
  \theta\to\theta+2\pi,
\end{equation}
with the KK levels relabeled by $n\to n\mp1$ for the two charge sectors. 

\paragraph{Orbifold compactification.}
Now we take the compact space to be the interval $0\leq y\leq \ell$, obtained from a covering
circle by the reflection $y\to-y$.  We use the same gauge twist as in the main text. For a fundamental doublet, the adjoint action can be lifted to the matrix
$P_\Psi=\sigma^1$.  The standard spinor reflection on an
$S^1/\Z_2$ orbifold is then
\begin{equation}
  \Psi(x,-y)=\eta_0\,\gamma_5 P_\Psi\Psi(x,y),
  \label{eq:fermion-orbifold-y0}
\end{equation}
at the fixed plane $y=0$, and
\begin{equation}
  \Psi(x,2\ell-y)=\eta_\ell\,\gamma_5 P_\Psi\Psi(x,y),
  \label{eq:fermion-orbifold-yell}
\end{equation}
at the fixed plane $y=\ell$, where
\begin{equation}
  \eta_0,\eta_\ell=\pm1
\end{equation}
are intrinsic parities and $\gamma_5$ is the four-dimensional chirality matrix. Orbifold fermion boundary conditions of this form are standard in five-dimensional
interval compactifications and orbifold gauge theories, see e.g.~\cite{Kawamura:2000ev,Hall:2001pg,ArkaniHamed:2001is,Scrucca:2001eb}.

At either fixed plane $y_i=0,\ell$, define the internal projectors
\begin{equation}
  \Pi_i^\pm={1\over2}\left(1\pm\eta_iP_\Psi\right),
  \qquad \eta_i=\eta_0,\eta_\ell .
\end{equation}
Equations~\eqref{eq:fermion-orbifold-y0} and \eqref{eq:fermion-orbifold-yell} are
equivalent to the boundary conditions
\begin{equation}
  \Pi_i^-\Psi_R(x,y_i)=0,
  \qquad
  \Pi_i^+\Psi_L(x,y_i)=0 .
  \label{eq:fermion-fixed-plane-projectors}
\end{equation}
Thus the right-handed component lives in the $+\eta_i$ eigenspace of $P_\Psi$, while the
left-handed component lives in the $-\eta_i$ eigenspace.  In the Cartan basis,
$P_\Psi=\sigma^1$ exchanges the two $T^3$-charge components,
\begin{equation}
  \psi_+(x,-y)=\eta_0\gamma_5\psi_-(x,y),
  \qquad
  \psi_-(x,-y)=\eta_0\gamma_5\psi_+(x,y),
  \label{eq:cartan-components-reflection}
\end{equation}
and similarly at $y=\ell$.  Therefore the $q=+1$ and $q=-1$ components are not two
independent interval fields with independently chosen boundary conditions, but two pieces of one
orbifold-projected bulk fermion.

To see the Wilson-line dependence, consider a constant background
\begin{equation}
  \theta_I(x)
  \equiv
  \int_0^\ell dy\,\mathcal A_5(x,y).
\end{equation}
As on the circle, we may gauge away the constant $\mathcal A_5$ in the interior, in which case the Wilson line then appears as a relative rotation between the boundary
conditions at $y=0$ and $y=\ell$.  Solving the resulting one-dimensional
eigenvalue problem gives a projected tower that may be labeled as
\begin{equation}
  p_{5,n}
  =
  \frac{\pi(n+\alpha)+\theta_I}{\ell},
  \qquad
  n\in\Z,
  \label{eq:orbifold-projected-momenta}
\end{equation}
where
\begin{equation}
  \alpha=
  \begin{cases}
    0, & \eta_0\eta_\ell=+1,\\[1mm]
    \frac12, & \eta_0\eta_\ell=-1.
  \end{cases}
  \label{eq:orbifold-alpha}
\end{equation}

This is a single projected tower, in the sense that the $q=-1$, $m=-m_\Psi$ Cartan component is the
Weyl-reflected partner inside the same interval eigenfunction, not a second independent
tower.

Finally, note that a constant five-dimensional Dirac mass $M\bar\Psi\Psi$
is odd under the orbifold reflection and is not allowed as a constant bulk term. On $S^1/\Z_2$ the analogous
allowed bulk mass is a kink mass $M(y)\bar\Psi\Psi$ satisfying $M(-y)=-M(y)$.
In contrast, the adjoint Yukawa coupling
\begin{equation}
  -y_5\bar\Psi\Phi^a\sigma^a\Psi
\end{equation}
is allowed precisely because the adjoint scalar has the extra intrinsic minus sign.

\section{Axion--Yang Mills on the interval \label{app:trueCS}}

As noted in Sec.~\ref{sec:orbifold-global}, there is an apparent puzzle regarding the quantization of the axion--Yang Mills coupling on the interval: a single bulk fermion can generate an odd local parity-anomaly
density in the interior of the interval, which leads to a half-integer coefficient for the $2\pi$-periodic orbifold axion
$\vartheta$. The resolution is that the
local interval Chern--Simons density is not the complete parity-odd phase
of the orbifold fermion determinant.  The complete determinant also knows
about the boundary conditions that exchange the two Cartan components at
the fixed planes.
 
A useful way to organize the argument is to begin with the covering-circle
theory.  Before the orbifold projection, the two Cartan components have
\begin{equation}
  (q_+,m_+)=(+1,+m_\Psi),
  \qquad
  (q_-,m_-)=(-1,-m_\Psi).
\end{equation}
Each contributes
\begin{equation}
  \Delta k_\pm
  =
  -\frac12 q_\pm\,\sgn(m_\pm)
\end{equation}
to the mixed Chern--Simons coefficient, so the two contributions add:
\begin{equation}
  k_{\rm cover}
  =
  -\sgn(m_\Psi).
  \label{eq:cover-cs-level-app}
\end{equation}
For a reflection-symmetric zero-mode background, the Wilson line around
the covering circle is
\begin{equation}
  \theta_{\rm cover}
  =
  \oint_{S^1_{\rm cover}}\mathcal A_5dy
  =
  2\int_0^\ell\mathcal A_5dy
  =
  \vartheta .
  \label{eq:cover-holonomy-vartheta}
\end{equation}
The orbifold projection necessarily reorganizes this between a local contribution in the
interior and a contribution fixed by the orbifold boundary conditions.

This can be robustly justified in terms of the boundary $\eta$-invariant, or more prosaically from the KK determinant.  Put the
four-dimensional $SU(N)$ zero mode in a background with instanton number
\begin{equation}
  Q_N
  =
  \frac{1}{8\pi^2}
  \int_{M_4}\tr_{\bf N}G\wedge G
  \in\Z .
  \label{eq:QN-threshold}
\end{equation}
As discussed in Appendix \ref{app:fermion-bcs}, the orbifold reflection
exchanges the $q=+1$ and $q=-1$ Cartan components.  A physical interval
mode therefore contains both components and is described by the single
projected tower
\begin{equation}
  p_{5,n}(\theta_I)
  =
  \frac{\pi(n+\alpha)+\theta_I}{\ell}.
\end{equation}
At each KK level, the four-dimensional Dirac operator contains the complex
mass
\begin{equation}
  m_\Psi+i\gamma_5p_{5,n}.
\end{equation}
A four-dimensional Dirac fermion in the fundamental representation with
complex mass $m+i\gamma_5p$ shifts the $SU(N)$ theta angle by the anomalous
chiral rotation needed to make its mass real.  With the sign convention
chosen to match Eq.~\eqref{eq:fermion-shift}, the unregulated topological
phase is therefore
\begin{equation}
  \exp(i\Gamma)
  =
  \exp\left[
    -iQ_N\sum_{n\in\Z}
    \arg\left(m_\Psi+ip_{5,n}(\theta_I)\right)
  \right].
  \label{eq:unregulated-topological-phase}
\end{equation}
The sum over phases is not meaningful until the infinite tower is
regulated.

For the purpose of extracting the Wilson-line dependence, introduce a
Pauli--Villars field with the same gauge quantum numbers and the same
orbifold boundary conditions.  Its mass must be introduced in a way
compatible with the microscopic $SU(2)$ theory; after Higgsing, its two
Cartan components therefore also acquire opposite signed masses.  Denote
the corresponding projected mass parameter by $M_{\rm PV}$.  The
Wilson-line-dependent phase of the regulated determinant may be written as
\begin{equation}
  Z_{\Psi,{\rm top}}^{\rm reg}(\theta_I,Q_N)
  =
  \left[
  \frac{
    {\mathfrak D}_\alpha(m_\Psi,\theta_I)/
    |{\mathfrak D}_\alpha(m_\Psi,\theta_I)|
  }{
    {\mathfrak D}_\alpha(M_{\rm PV},\theta_I)/
    |{\mathfrak D}_\alpha(M_{\rm PV},\theta_I)|
  }
  \right]^{-Q_N},
  \label{eq:regulated-determinant-ratio-threshold}
\end{equation}
where
\begin{equation}
  {\mathfrak D}_\alpha(m,\theta_I)
  \equiv
  \prod_{n\in\Z}
  \left[
    m+\frac{i}{\ell}
    \bigl(\pi(n+\alpha)+\theta_I\bigr)
  \right]_{\rm reg}
  \propto
  \sin\bigl(\theta_I+\pi\alpha-im\ell\bigr).
  \label{eq:Dalpha-definition-threshold}
\end{equation}
The proportionality factor is independent of $\theta_I$ and does not
affect the axion coupling.

Equation~\eqref{eq:Dalpha-definition-threshold} also makes the orbifold
large gauge transformation transparent:
\begin{equation}
  {\mathfrak D}_\alpha(m,\theta_I+\pi)
  =
  -{\mathfrak D}_\alpha(m,\theta_I).
  \label{eq:Dalpha-large-shift}
\end{equation}
In an instanton sector of charge $Q_N$, the phase of the physical
determinant changes by $(-1)^{Q_N}$.  The regulator determinant changes by
the same factor, so the complete regulated determinant is invariant.  This
is the KK manifestation of the fact that the local interval
Chern--Simons term by itself is not gauge invariant under the allowed
orbifold large gauge transformation.

For $|m|\ell\gg1$ and $0<x<\pi$,
\begin{equation}
  \arg\sin(x-im\ell)
  =
  \sgn(m)\,x
  -\frac{\pi}{2}\sgn(m)
  +\order(e^{-2|m|\ell}).
  \label{eq:arg-sin-large-mass}
\end{equation}
Using Eqs.~\eqref{eq:regulated-determinant-ratio-threshold} and
\eqref{eq:Dalpha-definition-threshold}, one obtains
\begin{equation}
  \Gamma^{\rm reg}
  =
  -\Bigl[
    \sgn(m_\Psi)-\sgn(M_{\rm PV})
  \Bigr]\theta_IQ_N
  +\Theta_N^{\rm const}Q_N
  +\order\!\left(
    e^{-2|m_\Psi|\ell},
    e^{-2|M_{\rm PV}|\ell}
  \right),
  \label{eq:threshold-thetaI-result}
\end{equation}
where $\Theta_N^{\rm const}$ is independent of the Wilson line and may be
absorbed into the microscopic $SU(N)$ theta angle.

The relative sign of the regulator is not an independent low-energy
choice, but is fixed by matching the projected theory to the chosen definition of the fermion determinant on the
covering circle.  We choose the ultraviolet prescription in which the
covering-circle bifundamental induces the integer level
\eqref{eq:cover-cs-level-app}.  In the interval representation this
corresponds to
\begin{equation}
  \sgn(M_{\rm PV})=-\sgn(m_\Psi).
  \label{eq:PV-choice-threshold}
\end{equation}
Equation~\eqref{eq:threshold-thetaI-result} then gives
\begin{equation}
  \Gamma^{\rm reg}
  =
  -2\sgn(m_\Psi)\theta_IQ_N
  +\Theta_N^{\rm const}Q_N
  +\order(e^{-2|m_\Psi|\ell}),
\end{equation}
or, using $\vartheta=2\theta_I$,
\begin{equation}
  S_{\vartheta GG}^{(1\Psi)}
  =
  -\sgn(y_5v_5)
  \frac{1}{8\pi^2}
  \int_{M_4}
  \vartheta\,\tr_{\bf N}G\wedge G .
  \label{eq:one-fermion-orbifold-final-coupling}
\end{equation}
In terms of the canonically normalized axion
$a=f_{\rm orb}\vartheta$,
\begin{equation}
  \mathcal L_{4d}
  \supset
  -\sgn(y_5v_5)
  \frac{a}{32\pi^2f_{\rm orb}}
  G^A_{\mu\nu}\widetilde G^{A\mu\nu}.
  \label{eq:one-fermion-orbifold-components}
\end{equation}

\section{Radion EFT}
\label{app:radion-eft}

This appendix presents the radion EFT underlying the fixed-interval
approximation used in Sec.~\ref{sec:string-cosmology}.  In the main text the
interval is already stabilized and its proper length is denoted by $\ell$.
We keep that notation here and allow a fluctuating physical length
$\ell_{\rm phys}(x)=\ell[1+\delta(x)]$.  The approximation
requires the cosmological background to be four-dimensional,
$H_{\rm end}\ell\ll1$, and the thermal displacement of the stabilized
interval to be small.  A radion at the KK scale is sufficient: the conservative
benchmark is $H_{\rm end}\ll m_r\sim\ell^{-1}$, with any parametrically
heavier stabilization dynamics treated as UV data matched onto the
fixed-interval EFT.

We start from the five-dimensional action
\begin{equation}
  S
  =
  \int \dd^4x\,\dd y\,\sqrt{|G|}
  \left[
    \frac{M_5^3}{2}R_5
    +
    \cL_{5d}^{\rm bulk}
  \right]
  +
  S_{\rm stab}.
  \label{eq:cosmology-master-action}
\end{equation}
Here $G_{MN}$ is the five-dimensional metric, $G$ its determinant, $R_5$ its
Ricci scalar, and $M_5$ the five-dimensional Planck scale.  The bulk
Lagrangian $\cL_{5d}^{\rm bulk}$ contains the gauge and matter fields
discussed above, while $S_{\rm stab}$ denotes the bulk and/or fixed-plane
physics that stabilizes the interval size.

For the flat $S^1/\Z_2$ orbifold, with $0\leq y\leq\ell$ on the
fundamental interval, a convenient zero-mode metric ansatz is the
$A_\mu=0$ specialization of Ref.~\cite{Ponton:2001hq}, converted to our
mostly-minus convention,
\begin{equation}
  \dd s_5^2=
  \varphi^{-1/3}(x)g_{\mu\nu}(x)\,\dd x^\mu\dd x^\nu
  -\varphi^{2/3}(x)\dd y^2 .
  \label{eq:radion-metric}
\end{equation}
The Weyl factor puts $g_{\mu\nu}$ in four-dimensional Einstein frame.  The
proper interval length is therefore
\begin{equation}
  \ell_{\rm phys}(x)=\ell\varphi^{1/3}(x),
  \qquad
  \delta(x)\equiv\frac{\ell_{\rm phys}(x)-\ell}{\ell},
  \qquad
  \varphi=(1+\delta)^3 .
  \label{eq:radion-varphi-map}
\end{equation}
Reducing the five-dimensional Einstein-Hilbert term with
$\langle\varphi\rangle=1$ gives
\begin{equation}
  \frac{M_5^3}{2}\int \dd^5x\,\sqrt{|G|}\,R(G)
  =
  \frac{M_{\rm Pl}^2}{2}\int \dd^4x\,\sqrt{-g}\,
  \left[
    R_4(g)
    +
    \frac16
    \frac{g^{\mu\nu}\partial_\mu\varphi\partial_\nu\varphi}{\varphi^2}
  \right],
  \qquad
  M_{\rm Pl}^2=M_5^3\ell .
  \label{eq:radion-eh-reduction}
\end{equation}
Using Eq.~\eqref{eq:radion-varphi-map}, the linearized kinetic term is
\begin{equation}
  \cL_{\rm kin}\simeq
  \frac12\partial_\mu r\,\partial^\mu r,
  \qquad
  r\simeq\sqrt{\frac32}M_{\rm Pl}\delta .
  \label{eq:radion-linear-normalization}
\end{equation}
Thus $r$ is the canonically normalized radion fluctuation at linear order,
while $\delta$ is the fractional displacement of the physical interval
length.

We do not need to specify the microscopic stabilization mechanism.  Bulk,
brane, Casimir, and threshold effects are encoded in a renormalized
zero-temperature potential $V_{\rm rad}$.  A microscopic construction would
impose the full matching and tuning conditions, such as those in
Ref.~\cite{Kanti:2002fx}; the local EFT only needs
\begin{equation}
  V_{\rm rad}(\ell)=0,
  \qquad
  \left.\frac{\partial V_{\rm rad}}{\partial\ell_{\rm phys}}\right|_{\ell}=0,
  \qquad
  \left.\frac{\partial^2 V_{\rm rad}}{\partial\ell_{\rm phys}^2}\right|_{\ell}>0 .
  \label{eq:radion-stability-conditions}
\end{equation}
The additive constant tunes the four-dimensional cosmological constant, the
first derivative fixes the interval size, and the second derivative
stabilizes the radion.  We define $m_r$ by the local expansion
\begin{equation}
  V_{\rm rad}
  =
  \frac12m_r^2r^2+\cdots
  =
  \frac34M_{\rm Pl}^2m_r^2\delta^2+\cdots,
  \qquad
  \frac32M_{\rm Pl}^2m_r^2
  =
  \ell^2
  \left.\frac{\partial^2 V_{\rm rad}}{\partial\ell_{\rm phys}^2}\right|_{\ell}.
  \label{eq:radion-appendix-potential}
\end{equation}

The thermal force on the radion follows directly from the stress tensor.  For
matter described by a five-dimensional stress tensor $T^{MN}$,
\begin{equation}
  \delta S_{\rm matter}
  =
  -\frac12\int\dd^5X\,\sqrt{|G|}\,T^{MN}\delta G_{MN}.
  \label{eq:radion-stress-variation}
\end{equation}
For an isotropic relativistic fluid,
\begin{equation}
  T^M_{\phantom{M}N}
  =
  {\rm diag}(\rho_{5d},-P_5,-P_5,-P_5,-P_5),
  \label{eq:radion-thermal-stress-tensor}
\end{equation}
so the component conjugate to a change of the compact direction is the
pressure $-T^5_{\phantom{5}5}=P_5$.  It is simplest to extract this force at
fixed induced four-dimensional metric, before the Weyl transformation to
Einstein frame.  Around the stabilized background, Eq.~\eqref{eq:radion-metric}
gives
\begin{equation}
  \delta G_{55}
  =
  2G_{55}\frac{\delta\ell_{\rm phys}}{\ell_{\rm phys}},
\end{equation}
which gives the linearized local work term
\begin{equation}
  \delta S_{\rm th}
  =
  \int\dd^4x\,\sqrt{-g}\,P_5\,\delta\ell_{\rm phys}.
  \label{eq:radion-thermal-work}
\end{equation}
Equivalently, the thermal contribution to the four-dimensional potential
satisfies
\begin{equation}
  \left.
  \frac{\partial V_{\rm th}}{\partial\ell_{\rm phys}}
  \right|_T
  =
  -P_5 .
  \label{eq:radion-thermal-force}
\end{equation}
Combining this with Eq.~\eqref{eq:radion-appendix-potential}, the displaced
finite-temperature minimum obeys
\begin{equation}
  \frac{3M_{\rm Pl}^2m_r^2}{2\ell}\delta
  =
  P_5.
  \label{eq:radion-thermal-balance}
\end{equation}
For a locally five-dimensional relativistic plasma, using the same
normalization convention for $N_{\rm th}^{(5)}$ as in Sec.~\ref{sec:string-cosmology},
\begin{equation}
  P_5
  =
  \frac14\rho_{5d}
  =
  \frac{N_{\rm th}^{(5)}}{4}
  \int\frac{\dd^4p}{(2\pi)^4}
  \frac{|p|}{e^{|p|/T}-1}
  =
  \frac{3\zeta(5)}{4\pi^2}\,N_{\rm th}^{(5)}T^5 .
  \label{eq:radion-pressure-expression}
\end{equation}
This gives the finite-temperature displacement that controls the
fixed-interval approximation,
\begin{equation}
  \frac{\Delta\ell}{\ell}
  =
  \delta
  \simeq
  \frac{\zeta(5)N_{\rm th}^{(5)}\ell T^5}
  {2\pi^2M_{\rm Pl}^2m_r^2}.
  \label{eq:radion-temperature-displacement}
\end{equation}
The fixed-interval treatment requires this shift to be small at the largest
temperature reached during reheating.  With the instantaneous-reheating
estimate used in Sec.~\ref{sec:string-cosmology}, the four-dimensional
Friedmann relation gives
\begin{equation}
  \frac{\Delta\ell}{\ell}
  \simeq
  \frac12
  \left(\frac{H}{m_r}\right)^2 .
  \label{eq:radion-displacement-hubble}
\end{equation}
Thus a radion heavy compared with the Hubble scale has negligible displacement
throughout the post-inflationary history.

\bibliographystyle{JHEP}
\bibliography{axionstring}

\end{document}